\documentclass[twocolumn,aps,pre,10pt,floatfix,longbibliography]{revtex4-2}
\usepackage[switch,columnwise]{lineno}
\usepackage{lineno} 
\usepackage{graphicx}% Include figure files
\usepackage{dcolumn}% Align table columns on a decimal point
\usepackage{bm}% bold math
\usepackage{setspace}
\usepackage{fancyhdr}
\usepackage{amsmath}
\usepackage{upgreek}
\usepackage{amssymb}
\usepackage{amsfonts}
\usepackage{textcomp, gensymb}
\usepackage{color}
\usepackage{epstopdf}
\usepackage{booktabs}
\usepackage{relsize}
\usepackage{xcolor}
\usepackage{natbib}
\usepackage{float}
\newcommand{\volume}{{\ooalign{\hfil$V$\hfil\cr\kern0.08em--\hfil\cr}}}
\usepackage{setspace}
\usepackage[hidelinks]{hyperref}
\hypersetup{
     colorlinks=true,
     linkcolor=blue,
     filecolor=blue,
     citecolor = blue,      
     urlcolor=blue,
     }
\usepackage{multirow}
\allowdisplaybreaks

\def\rtx@apspre{%
\class@info{APS journal PRE selected}%
}%

\begin{document}
 \title{Color-gradient lattice Boltzmann modeling of wetting boundary condition on curved solid boundaries}
\author{M. Bhattacharya$^1$}
\author{S. Dash$^2$}
\author{M. Sutar$^3$}
\author{J. Ravinder$^3$}
\author{M. Nimalan$^3$}
\author{A. Subhedar$^1$}
\email[Email address for correspondence: ]{amol.subhedar@iitb.ac.in}
\affiliation{$^{1}$Department of Chemical Engineering, Indian Institute of Technology Bombay, Maharashtra-400076, India}
\affiliation{$^{2}$Department of Chemical Engineering, Indian Institute of Technology Delhi, Delhi-110016, India}
\affiliation{$^{3}$Engineering for Research (e4r), Thoughtworks, Pune, India}

\date{\today}
%%%%%%%%%%%%%%%%%%%%%%%%%%%%%%%%%%%%%%%%%%%%%%%%%%%%%%%%%%%%%%%%%%%%%%%%%%%%%%%%
\begin{abstract}
We introduce a wetting boundary condition for curved solid boundaries within a diffuse interface framework for lattice Boltzmann method. The boundary condition relies on updating the order parameter (color/phase-field) values on ghost nodes inside the solid phase. The ghost node color modification rule, in turn, extends the equilibrium color profile into the solid phase.  Numerical simulations performed on an NVIDIA A100 GPU demonstrate that the wetting scheme retains the model's ability to handle large density and viscosity contrasts while producing relatively small spurious currents.  The present scheme agrees well with analytical solutions/other numerical works for both static and dynamic contact lines on curved solid boundaries.
\end{abstract}
\maketitle

\section{Introduction}

Wetting phenomena play an instrumental role in many technologically important processes, spanning multiple orders of magnitude in length and time scales, such as porous media flow~\cite{parker1989multiphase}, colloidal film drying~\cite{routh2013drying}, curtain coating~\cite{kistler1994teapot} and microfluidics~\cite{utada2007dripping}. At low capillary and Weber numbers, surface tension effects primarily control the bulk dynamics by introducing a pressure difference across the liquid-gas interface. In addition, surface tension effects can influence the system dynamics via the wetting boundary condition even at high Reynolds and Weber numbers ~\cite{bergeron2000controlling,song2022droplet}. The wetting boundary condition models a contact line moving with a constant contact angle or maintaining a contact angle within a given (hysteresis) window.   In this work, we focus on modeling the former for immiscible liquid-gas systems using lattice Boltzmann method. The lattice Boltzmann method~\cite{aidun2010lattice,kruger2017lattice} has emerged as a reliable and efficient tool for numerical investigation of such systems, given its parallelizability and ability to handle complex boundary conditions with relative ease. 

Accurate implementation of the wetting boundary condition is central to maintaining the overall reliability of the numerical simulations involving contact lines. For example, accuracy of the wetting boundary condition directly influences errors in the measurement of capillary forces on solid particles~\cite{liu2015diffuse,bhattacharya2026Capillary}. In continuum models, the wetting boundary condition is stated for sharp interfaces that relates the angle between the normal to the solid boundary and the normal to the liquid-gas interface to the equilibrium contact angle. On the other hand, common numerical frameworks~\cite{shams2018numerical} represent the fluid–fluid interface as diffuse, while the solid-fluid boundary remains sharp. An order parameter (phase-field, density or color-field) is used to distinguish between fluid phases. Diffuse interface models strive to mimic the sharp interface wetting boundary condition locally across the diffuse interface by adjusting the gradient of the order parameter at the contact line. In this context, it is useful to introduce boundary nodes for fluid and solid phases. The solid (fluid) boundary nodes are those with at least one neighbor that is fluid (solid), as defined by the given stencil. The solid boundary nodes are also referred to as ghost nodes, as fluid dynamical quantities are not evaluated at these nodes.

Numerous approaches have been employed to model wetting boundary conditions in diffuse interface frameworks. These include: prescribing a constant order parameter at ghost nodes (fictitious density methods)~\cite{martys1996simulation,li2014contact}, reorienting the unit normal to the phase field at fluid boundary nodes~\cite{leclaire2017generalized,akai2018wetting},  geometrically enforcing the contact angle at the interface~\cite{ding2007wetting,zhang2023simplified,wang2024wetting},  incorporating solid–fluid surface energy formulations~\cite{lee2008wall}, and linear extrapolation of the order parameter at the solid boundary~\cite{fakhari2017diffuse}. These approaches may suffer from several limitations including: introduction of artificial mass source or sink at solid boundaries~\cite{yiotis2007lattice}, generation of spurious velocities near interfaces~\cite{connington2012review, ezzatneshan2020evaluation}, restrictions to planar solid geometries~\cite{liu2015lattice, liang2019lattice}, increased non-locality due to the use of higher-order stencils for gradient evaluation~\cite{liu2015lattice}, inconsistencies between the locations where no-slip and wetting boundary conditions are enforced~\cite{leclaire2017generalized}, and errors introduced due to linearization of the order parameter profile near solid boundary~\cite{fakhari2017diffuse}.

We introduce a wetting boundary condition for curved solid boundaries in two dimensions. The scheme relies on adjusting the order parameter on  ghost nodes such that the equilibrium order parameter profile is extended in the solid phase.  A color-gradient model~\cite{subhedar2020interface, subhedar2022color} that employs a velocity-based equilibrium distribution function is used to solve fluid dynamical equations for two immiscible fluid phases.  Accuracy of the proposed scheme is assessed against analytical solutions and the scheme of Fakhari \textit{et al.}~\cite{fakhari2017diffuse} for both static and dynamic contact lines on curved solid boundaries. Numerical simulations are performed using a JAX-based~\cite{jax2018github} framework running on NVIDIA A100 GPUs. We also quantify spurious currents near contact lines and assess the computational performance of the implementation on GPUs.

The rest of the article is organized as follows: Fluid dynamical equations in the bulk phases and the proposed wetting boundary condition are introduced in section~\ref{section:wetting}. Numerical benchmarks comparing the proposed scheme with analytical solutions and other numerical schemes, along with computational details, are elaborated in section~\ref{section:numericalResults}. A summary of current work is discussed in section~\ref{section:conclusion}.

\section{Wetting Boundary condition}\label{section:wetting}
We use a color-gradient model~\cite{subhedar2022color} to simulate incompressible immiscible Newtonian two-phase flow. The emergent equations of the model are
\begin{align}
    \nabla \cdot \mathbf{u} &= 0, \\
    \frac{\partial \mathbf{u}}{\partial t} + \mathbf{u}\cdot\nabla \mathbf{u} &= \frac{1}{\rho} \left( \nabla \cdot \bf{\Pi} + {\mathbf{F}_{\text{s}}} + \rho\mathbf{g} \right), \label{eq:momentumBalance}\\
   \frac{\partial \varphi}{\partial t} + \nabla \cdot (\varphi\mathbf{u}) &= \nabla \cdot\left[M_{\varphi}\left(\nabla \varphi -\frac{4\varphi(1-\varphi)}{W}\mathbf{\hat n}_{\varphi}\right)\right] \label{eq:phaseFieldEvolution}, 
\end{align}
where $\mathbf{u}$ is fluid velocity, $\rho$ is fluid density,  $\mathbf{\Pi} = -p_{\text{h}}\mathbf{I} + \mu\left(\nabla \mathbf{u} + \nabla \mathbf{u}^T\right)$ is stress tensor, $p_{\text{h}}$ is pressure, $\mathbf{I} $ is identity matrix, $\mathbf{F}_{\text{s}}$ is surface tension force, $\mathbf{g}$ is acceleration due to gravity, $\varphi$ is color-field (phase-field) indicator that differentiates fluid phases, $W$ is liquid-gas interface width, $M_{\varphi}$ is liquid-gas interface mobility and $\mathbf{\hat n}_{\varphi} = \frac{\nabla \varphi}{|\nabla \varphi|}$ is unit normal to the liquid-gas interface defined when $|\nabla \varphi| \ne 0$. The color-field takes value $\varphi =1$ in the liquid (heavier) and $\varphi = 0$ in the gas (lighter) phase. We use subscripts l and g to denote the liquid and gas phase properties, respectively. The fluid density and kinematic viscosity are defined as $\rho = \rho_{\text{l}}\varphi + \rho_{\text{g}}(1-\varphi)$ and $ \frac{1}{\nu} = \frac{\varphi}{\nu_{\text{l}}} +  \frac{1-\varphi}{\nu_{\text{g}}}$, respectively.
The surface tension force $\mathbf{F}_{\text{s}}$ in Eq.~(\ref{eq:momentumBalance}) is modeled as 
\begin{align}
    \mathbf{F}_{\text{s}} = -\sigma \left [ \frac{48 }{W} \varphi (1-\varphi )\left(\varphi - \frac{1}{2}\right) + \frac{3 W}{2} \nabla^2 \varphi \right ] \nabla \varphi,
    \label{eq:surfaceTensionForce}
\end{align}
where $\sigma$ is the liquid-gas surface tension. In the absence of fluid velocity $\mathbf{u}$, equilibrium color profile $\varphi^{\text{eq}}$, solution of Eq.~(\ref{eq:phaseFieldEvolution}), is given by
\begin{align}
    \varphi^{\text{eq}}(\eta) = \frac{1}{2}\left(1\pm \tanh\left[\frac{2 \eta}{W}\right]\right), \label{eq:equilibriumPhaseField}
\end{align}
where $\eta$ is a local curvilinear coordinate that is normal to the liquid-gas interface, such that $\eta = 0$ identifies the sharp interface. The equilibrium solution does not depend upon the coordinate that is perpendicular to the $\eta$ coordinate. A useful property of the equilibrium solution, Eq.~(\ref{eq:equilibriumPhaseField}), is that the derivative with respect to $\eta$ can be expressed locally as a function of $\varphi$ alone. 
\begin{align}
    \frac{\partial \varphi^{\text{eq}}}{\partial \eta} = \frac{4}{W}\varphi^{\text{eq}}(1-\varphi^{\text{eq}}). \label{eq:equilibriumProfile}
\end{align}

\begin{figure}[htbp]
    \centering
    \includegraphics[width=\linewidth]{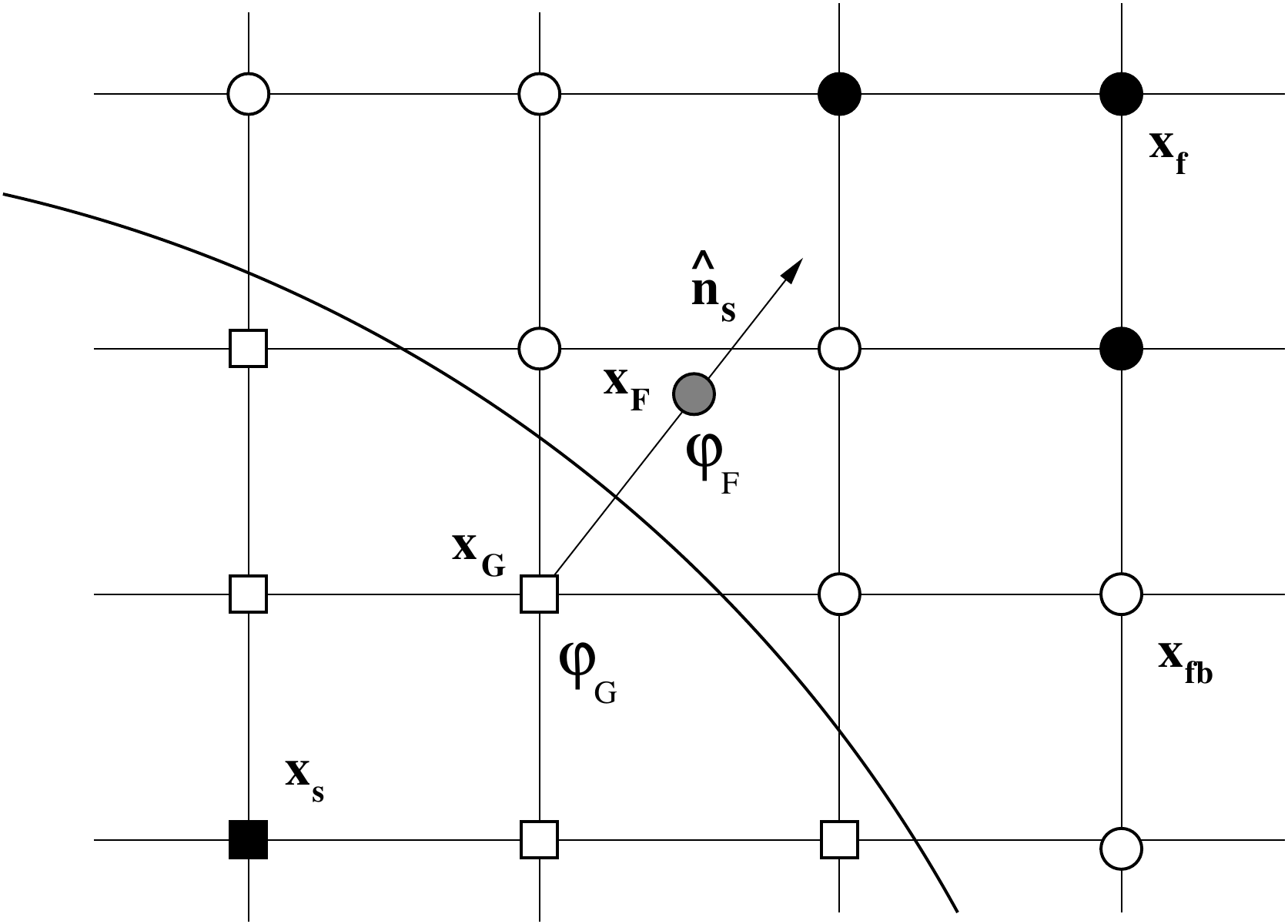}
    \caption{Schematic depicts a part of a solid boundary (solid black line) and computational nodes in the vicinity. Solid boundary (bulk) nodes are shown with hollow (filled) squares. Similarly, fluid boundary (bulk) nodes are shown with hollow (filled) circles. The wetting boundary condition, Eq.~(\ref{eq:ghost-nodePrescription}), is enforced by updating ghost node color-field $\varphi_{\text{G}}$ at $\mathbf{x}_{\text{G}}$ based upon the contact angle $\theta$ and color-field $\varphi_{\text{F}}$ at a point $\mathbf{x}_{\text{F}}$ in the fluid domain that is distance $\Delta x$ away from $\mathbf{x}_{\text{G}}$ in the direction of normal to the solid surface $\mathbf{\hat n}_{\text{s}}$.    }
    \label{fig:ghostNodeUpdate}
\end{figure}

\begin{figure}[htbp]
    \centering
    \includegraphics[width=0.8\linewidth]{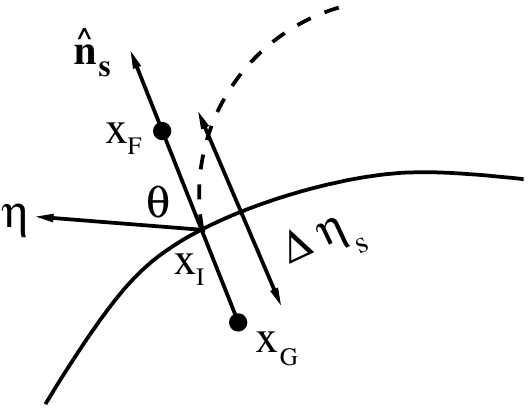}
    \caption{Ghost node $\mathbf{x}_{\text{G}}$ and corresponding point in the fluid domain $\mathbf{x}_{\text{F}}$ are related by $\mathbf{x}_{\text{F}} = \mathbf{x}_{\text{G}} + \Delta x \mathbf{\hat n}_{\text{s}}$ and the line joining them intersects the solid boundary (shown with black solid curved line) at $\mathbf{x}_{\text{I}}$. The liquid-gas interface is shown with black dashed line. We assume that the contact angle $\theta$ does not vary significantly along the line joining $\mathbf{x}_{\text{G}}$ and $\mathbf{x}_{\text{F}}$. In this case, the derivative of the color profile in the direction normal to the solid surface can be approximated as $\mathbf{\hat n}_{\text{s}}\cdot \nabla\varphi =\mathbf{\hat n}_{\text{s}}\cdot \mathbf{\hat n}_{\varphi} |\nabla\varphi| = -\cos\theta|\nabla \varphi| $ along the line joining $\mathbf{x}_{\text{G}}$ and $\mathbf{x}_{\text{F}}$.    }
    \label{fig:directionalDerivative}
\end{figure}

The wetting boundary condition is written as
\begin{align}
    \mathbf{\hat n}_{\text{s}}\cdot\mathbf{\hat n}_{\varphi} = -\cos\theta, \label{eq:sharpWettingCondition}
\end{align}
where $\mathbf{\hat n}_{\text{s}}$ is a unit normal to the solid boundary pointing away from the surface. The contact angle $\theta$ is the angle between the tangent to the solid surface and the tangent to the liquid-gas interface, measured from the liquid phase ($\varphi=1$).  The boundary condition is valid on the solid surface, which may not be aligned with the underlying rectangular grid, see Fig.~(\ref{fig:ghostNodeUpdate}). We denote the equal grid spacing in $x$ and $y$ directions as $\Delta x =\Delta y =1$.  The idea is to modify the color-field $\varphi_{\text{G}}$ at the ghost node $\mathbf{x}_{\text{G}}$, based on the color-field $\varphi_{\text{F}}$ at point $\mathbf{x}_{\text{F}}$ in the fluid domain, such that the wetting boundary condition is satisfied on the solid boundary. The point $\mathbf{x}_{\text{F}}$ is at a distance of $\Delta x$ from the ghost node $\mathbf{x}_{\text{G}}$ in the direction of outward normal to the solid boundary, see Fig.~(\ref{fig:directionalDerivative}). Thus, the ghost node $\mathbf{x}_{\text{G}}$ and fluid point $\mathbf{x}_{\text{F}}$ lie on opposite side of the solid boundary such that $\Delta \eta_{\text{s}} =\Delta x = (\mathbf{x}_{\text{F}}-\mathbf{x}_{\text{G}})\cdot\mathbf{\hat n}_{\text{s}} $, where $\Delta \eta_{\text{s}}$  is increment in normal to the solid boundary coordinate $\eta_{\text{s}}$. In general, $\mathbf{x}_{\text{F}}$ may not lie on any of the grid nodes. In that case, $\varphi_{\text{F}}$ is obtained from bilinear interpolation.

As the first step, we write the derivative of the equilibrium order parameter profile in the direction of the solid normal. Assuming color distribution is not far from equilibrium $\varphi \approx \varphi^{\text{eq}}$, color gradient is approximated as $\nabla \varphi \approx \nabla \varphi^{\text{eq}} = |\nabla \varphi^{\text{eq}}|\mathbf{\hat n}_{\varphi} = \frac{4}{W}\varphi(1-\varphi)\mathbf{\hat n}_{\varphi}$.   Using Eq.~(\ref{eq:equilibriumProfile}) in Eq.~(\ref{eq:sharpWettingCondition}) gives,
\begin{align}
    \frac{\partial \varphi}{\partial \eta_{\text{s}}} = -\frac{4}{W}\varphi(1-\varphi)\cos\theta, \label{eq:odePhiNs}
\end{align}
where $\frac{\partial \varphi}{\partial \eta_{\text{s}}} = \mathbf{\hat n}_{\text{s}}\cdot\nabla\varphi $ is color derivative in the direction normal to the solid surface. Equation~(\ref{eq:odePhiNs}) can also be derived from a surface energy argument~\cite{liang2019lattice}. Assuming the contact angle $\theta$ does not change significantly, integration of Eq.~(\ref{eq:odePhiNs}) from $\mathbf{x}_{\text{G}}$ to $\mathbf{x}_{\text{F}}$ gives
\begin{align}
    \varphi_{\text{G}} = \frac{\varphi_{\text{F}}}{\varphi_{\text{F}}+(1-\varphi_{\text{F}})~\exp\varepsilon},  \label{eq:ghost-nodePrescription}
\end{align}
where $\varepsilon = -\frac{4\Delta \eta_{\text{s}}}{W}\cos\theta = -\frac{4\Delta x}{W}\cos\theta$, $\varphi_{\text{G}}$ is color-field at ghost node, $\varphi_{\text{F}}$ is color-field at $\mathbf{x}_\text{F}$. The interface width is chosen such that $W \ge4\Delta x$ to maintain numerical stability, resulting $|\varepsilon|\le 1$.  Given that $\exp\varepsilon > 0$, the denominator of Eq.~(\ref{eq:ghost-nodePrescription}) is always greater than $\varphi_{\text{F}}$. Thus, the ghost node color-field lies between $0$ and $1$. Far away from the contact line,  color-field takes values $\varphi = \varphi_{\text{F}}= 1$ or $\varphi =\varphi_{\text{F}} = 0$. Equation~(\ref{eq:ghost-nodePrescription}) yields $\varphi_{\text{G}} = \varphi_{\text{F}}$ in this case, preventing the artificial mass transfer along the solid surface. More importantly, location of the solid boundary does not enter in the prescription Eq.~(\ref{eq:ghost-nodePrescription}).

It is instructive to compare the ghost nodes update rule, Eq.~(\ref{eq:ghost-nodePrescription}), with that of Fakhari \textit{et al.}~\cite{fakhari2017diffuse}. In the current notation
\begin{align}
    \varphi^{\text{Fakhari}}_{\text{G}} = \frac{2}{\varepsilon}\left(1+\frac{\varepsilon}{2} -\sqrt{\left(1+\frac{\varepsilon}{2}\right)^2 -2\varepsilon \varphi_{\text{F}}} \right) - \varphi_{\text{F}} \label{eq:ghostNodeFakhari},
\end{align}
where the ghost node $\mathbf{x}_{\text{G}}$ and the corresponding point $\mathbf{x}_{\text{F}}$ are assumed to be at equal distance from the solid boundary. For neutral contact angle $\theta=90\degree$, parameter $\varepsilon $ becomes zero, requiring special care in evaluating Eq.~(\ref{eq:ghostNodeFakhari}). Such a special care is not required for the present scheme, Eq.~(\ref{eq:ghost-nodePrescription}). Note that, the situation of $\theta=90\degree$ may also occur when the contact line moves in contact angle hysteresis mode~\cite{bhattacharya2026hysteresis}, where the neutral contact angle lies between advancing and receding contact angles. 

Both the current (Eq.~(\ref{eq:ghost-nodePrescription})) and the scheme of Fakhari \textit{et al.}~\cite{fakhari2017diffuse} (Eq.~(\ref{eq:ghostNodeFakhari})), can be expressed as a series in the small parameter $\varepsilon$ for a given value of color-field $\varphi_{\text{F}}$ in the fluid domain. The difference between these two series gives,
\begin{align}
    \varphi_{\text{G}} - \varphi^{\text{Fakhari}}_{\text{G}} &= \frac{\varepsilon ^3}{12} \left[ 1 -3 (1-\varphi_{\text{F}} ) \varphi_{\text{F}} \right]\varphi_{\text{F}}(1-\varphi_{\text{F}}) \nonumber\\
    &+ \mathcal{O}(\varepsilon^4). \label{eq:differenceSeriesPrescription}
\end{align}
Equation~(\ref{eq:differenceSeriesPrescription}) suggests that in the bulk fluid phases $\varphi_{\text{F}}(1-\varphi_{\text{F}}) = 0$, two prescriptions are the same.  Further, differences in the two prescriptions appear at third order in $\varepsilon$. For a given contact angle $\theta$,  limit of $\varepsilon=-\frac{4\Delta x}{W}\cos\theta \to 0$ corresponds to increasing interface grid resolution, and the two prescriptions should approach each other. Numerical simulations substantiate this proposition, as we show in benchmark~\ref{subsection:dropContactAngle}.

We remark that Eq.~(\ref{eq:ghostNodeFakhari}) is a special case of prescription given by Fakhari \textit{et al.}~\cite{fakhari2017diffuse}, where the solid boundary is assumed to lie halfway between the ghost node location $\mathbf{x}_{\text{G}}$ and the corresponding fluid point $\mathbf{x}_{\text{F}}$. Their general prescription is capable of accommodating the relative position of the solid boundary location with respect to $\mathbf{x}_{\text{G}}$ and $\mathbf{x}_{\text{F}}$. For consistency, the bounce-back rule for the fluid and order parameter (color/phase-field) populations must also be based on interpolation schemes. The use of such interpolated bounce-back rules may exacerbate issues with mass and momentum conservation~\cite{fei2024pore,tao2016investigation,he2022improved}. We, therefore, avoid comparing the prescription of Fakhari \textit{et al.}~\cite{fakhari2017diffuse} that relies on the exact location of the solid boundary with the current prescription Eq.~(\ref{eq:ghost-nodePrescription}).

\section{Numerical benchmarks and GPU performance}\label{section:numericalResults}

We test the proposed wetting scheme in Eq.~(\ref{eq:ghost-nodePrescription}) with Fakhari~\textit{et al.}'s~\cite{fakhari2017diffuse} scheme in Eq.~(\ref{eq:ghostNodeFakhari}) in this section. We define the unit solid normal $\mathbf{\hat n}_{\text{s}}$  analytically for the solid particles used in this work. When solid boundary shapes are not analytically known, smooth solid normals can be obtained with numerical stencils that preserve isotropy of the gradient operator up to sixth order~\cite{sbragaglia2007diffOperators, akai2018wetting}.  We report all quantities appearing in this section in lattice units. The color-field is initialized as $\varphi=(1+\cos\theta)/2$ at the ghost nodes. For the two-dimensional simulations, $\Delta z = 1$ is taken as the unit depth in the direction perpendicular to the plane of the paper.   Unless stated otherwise, the liquid-gas interface mobility $M_{\varphi}=0.02$, interface width $W=4$, surface tension $\sigma=10^{-2}$, liquid-gas density ratio $\frac{\rho_\text{l}}{\rho_\text{g}} = 1000$, dynamic viscosity ratio $\frac{\mu_\text{l}}{\mu_\text{g}} = 100$, gas density $\rho_\text{g} = 1$, dynamic viscosity $\mu_\text{g} = 0.1$ are fixed for all the simulations.

\subsection{Drop on  solid  plate}
\label{subsection:dropContactAngle}

Consider a liquid drop placed on a stationary horizontal plate under negligible gravity. For a given mass (area $A$) and contact angle $\theta$, the steady-state equilibrium shape of the drop is a circular arc that is independent of the viscosity and density ratios of the fluids. 
This benchmark allows to assess the accuracy of the wetting boundary condition for a simple case of a solid boundary with zero curvature and where the steady-state drop shape is known analytically. 

\begin{figure}[htbp]
    \centering
    \includegraphics[width=\linewidth]{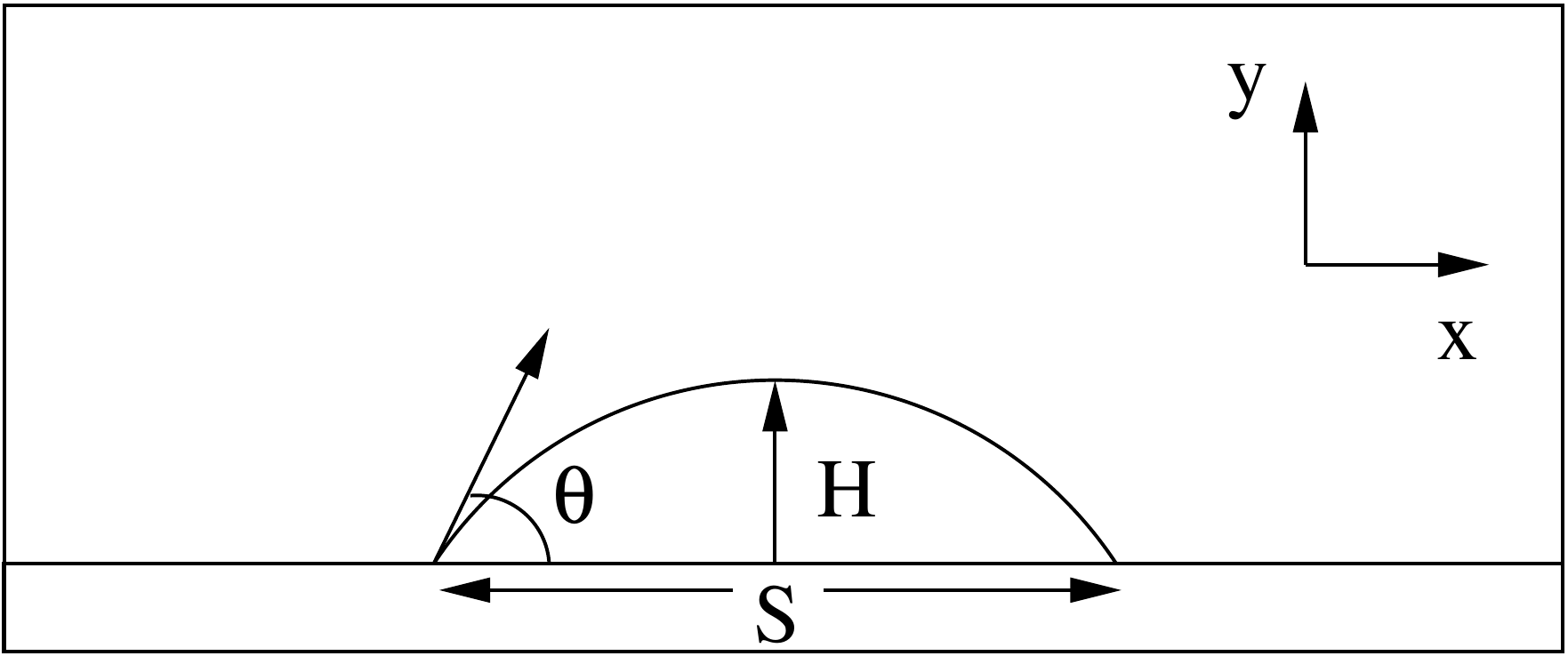}
    \caption{A liquid drop placed on a flat solid plate. For a given mass (area) of the drop, steady-state height $H$ and spread $S$ of the drop are used to estimate the resulting contact angle $\theta_{\text{m}}$ from the simulations.}
    \label{fig:dropPlateSchematic}
\end{figure}

\begin{figure*}[htbp]
    \centering
    \includegraphics[width=\linewidth]{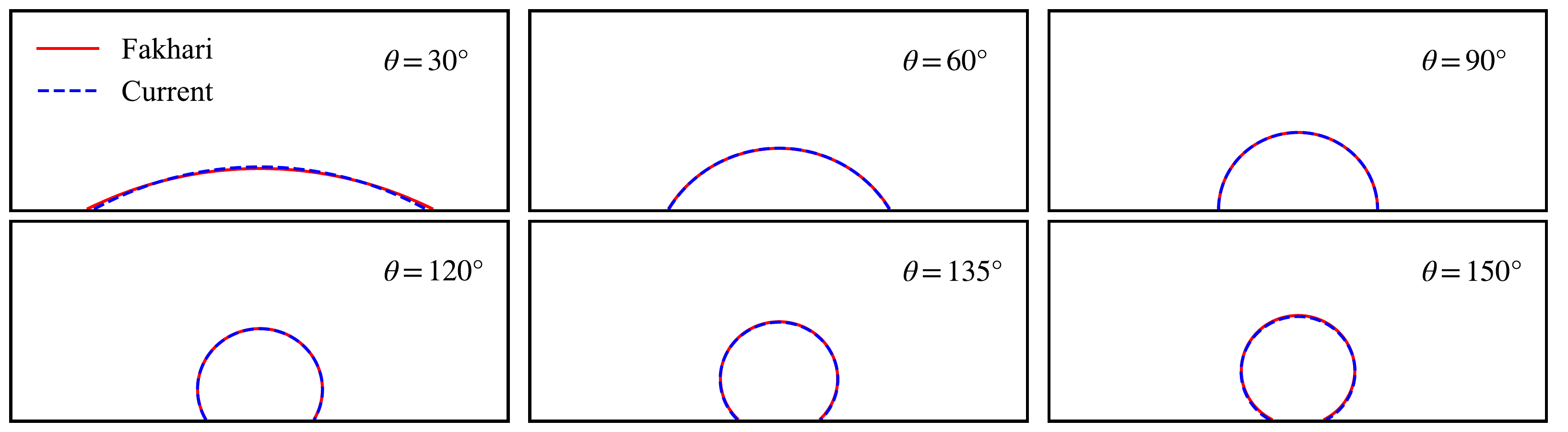}
    \caption{Steady-state shape of the drops for various contact angles $\theta$. The drop shape corresponds to contour of color-field $\varphi=\frac{1}{2}$. Solid red and dashed blue lines correspond to wetting boundary condition of Fakhari ~\textit{et al.}~\cite{fakhari2017diffuse} and present scheme, respectively.  Measured contact angles computed via Eq.~(\ref{eq:measuredContactAnglePlate}), and shown in Fig.~(\ref{fig:dropPlateErrorTheta}), highlight difference between the two wetting boundary conditions.}
    \label{fig:dropPlateContours}
\end{figure*}

\begin{figure}[htbp]
    \centering
    \includegraphics[width=0.9\linewidth]{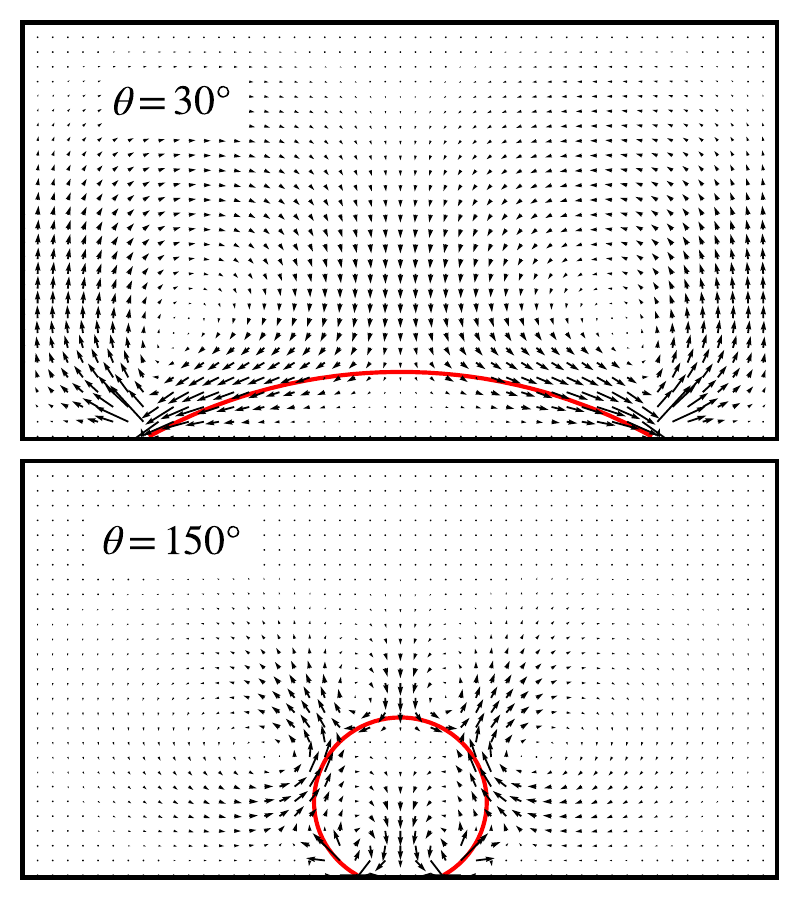}
    \caption{Spurious (parasitic) fluid velocity currents distribution for contact angles $\theta=30\degree$ and $\theta=150\degree$ obtained with the present wetting boundary condition. Magnitude of maximum fluid velocities for these two cases are $5.74 \times 10^{-7}$ and $7 \times 10^{-7}$, respectively.}
    \label{fig:dropPlateSpuriousCurrents}
\end{figure}

A schematic of the problem is shown in Fig.~(\ref{fig:dropPlateSchematic}). The capillary time scale $t_{\text{cap}} = \sqrt{\frac{\rho_{\text{l}} R_{\text{c}}^2 \Delta z}{\sigma}}$ and viscous time scale $t_{\text{visc}} = R^2_{\text{c}}/\nu_{\text{l}}$, where $R_{\text{c}}$ is equilibrium drop radius, dictate the typical time required to reach the equilibrium shape of the drop. The equilibrium drop radius $R_{c} = \sqrt{\frac{2A}{2\theta - {\sin 2\theta}}}$  continues to increase as $\theta$ decreases, thereby increasing the capillary time scale $t_{\text{cap}}$, especially for the high density of the drop. Thus, to reduce simulation time, we initialize the drop using the analytical shape for a given prescribed contact angle $\theta$ while ensuring less than $0.1\%$ error in the mass of the drop. For this benchmark, system size $(L_x,L_y) = (500,300)$ and area of the drop $A = 10^4$ are fixed. Boundary conditions are periodic in the $x$ direction and no-slip in the $y$ direction. Figure~(\ref{fig:dropPlateContours}) shows the steady-state drop shapes for different contact angles in the range of $30\degree\le\theta\le150\degree$. 
Corresponding spurious (parasitic) velocity distributions near the liquid-gas interface are shown in Fig.~(\ref{fig:dropPlateSpuriousCurrents}) for two contact angles $\theta= 30\degree$ and $\theta = 150\degree$. The maximum magnitude of the spurious currents for these two contact angles are $5.74 \times 10^{-7}$ and $7 \times 10^{-7}$, respectively for the proposed wetting scheme. For the scheme of Fakhari \textit{et al.}~\cite{fakhari2017diffuse}, the maximum magnitude of spurious currents for $\theta = 30\degree$ and $150\degree$ are $6.98 \times 10^{-7}$ and $1 \times 10^{-6}$ respectively.

\begin{figure}[htbp]
    \centering
    \includegraphics[width=\linewidth]{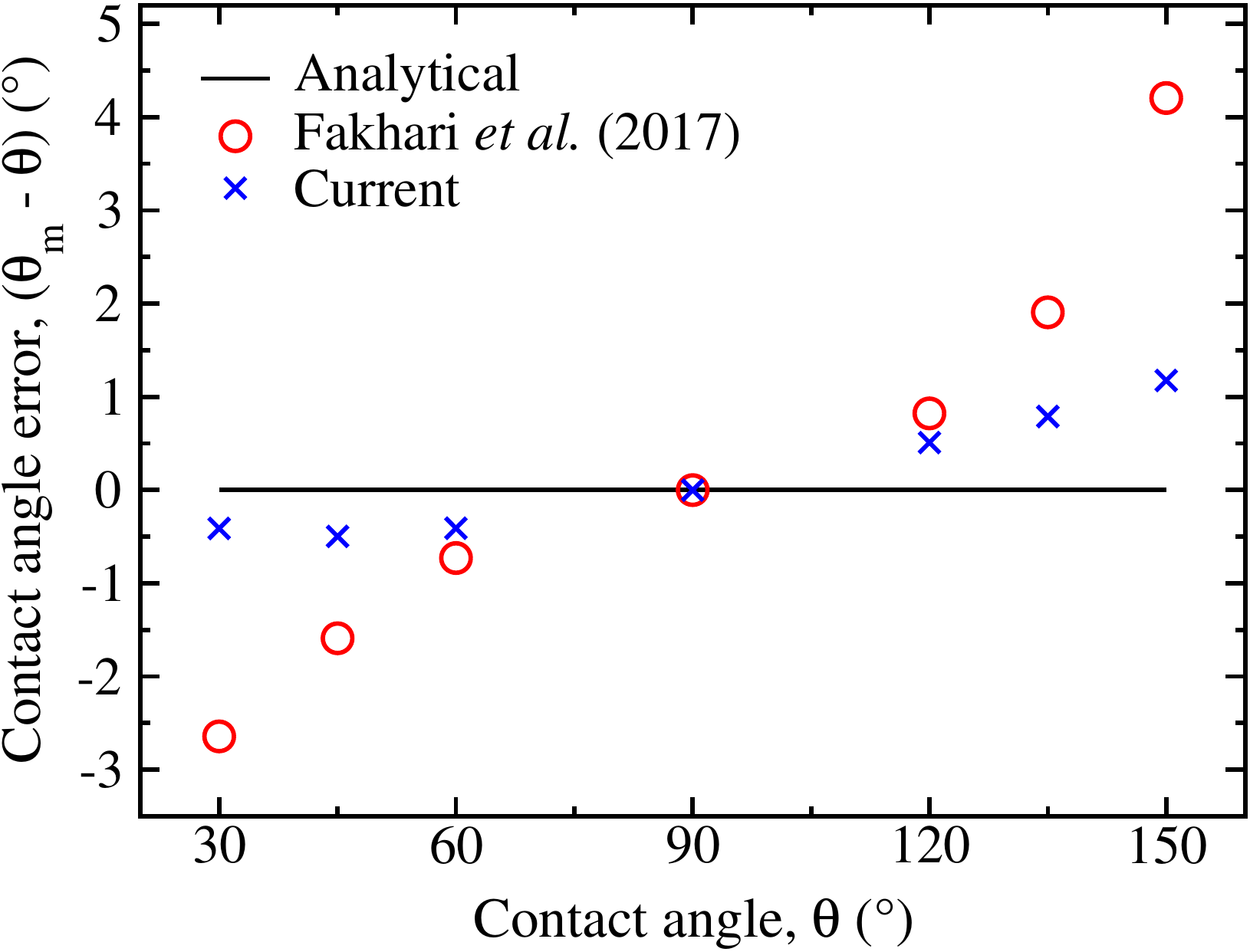}
    \caption{Difference of measured and set contact angle $\theta_{\text{m}} -\theta_{\text{}}$ as a function of set contact angle $\theta_{\text{set}}$. As the set contact angle deviates from the neutral one, the present wetting boundary scheme is more accurate than that of Fakhari~\textit{et al.}~\cite{fakhari2017diffuse}.  }
    \label{fig:dropPlateErrorTheta}
\end{figure}

Figure~(\ref{fig:dropPlateErrorTheta}) compares the prescribed and measured contact angles for interface width $W=4$. The contact angle is measured as
\begin{align}
\theta_{\text{m}} &= \arccos\left(\frac{S^2-H^2}{S^2 + H^2}\right), \label{eq:measuredContactAnglePlate}
\end{align}
where $S$ is the spread of the drop, $H$ is the height of the drop,  $\theta_{\text{m}}$ is measured contact angle, respectively. The proposed wetting boundary scheme, Eq.~(\ref{eq:ghost-nodePrescription}), does consistently better over that of Fakhari \textit{et al}~\cite{fakhari2017diffuse} for the range of prescribed contact angles.

\begin{figure}[htbp]
    \centering
    \includegraphics[width=\linewidth]{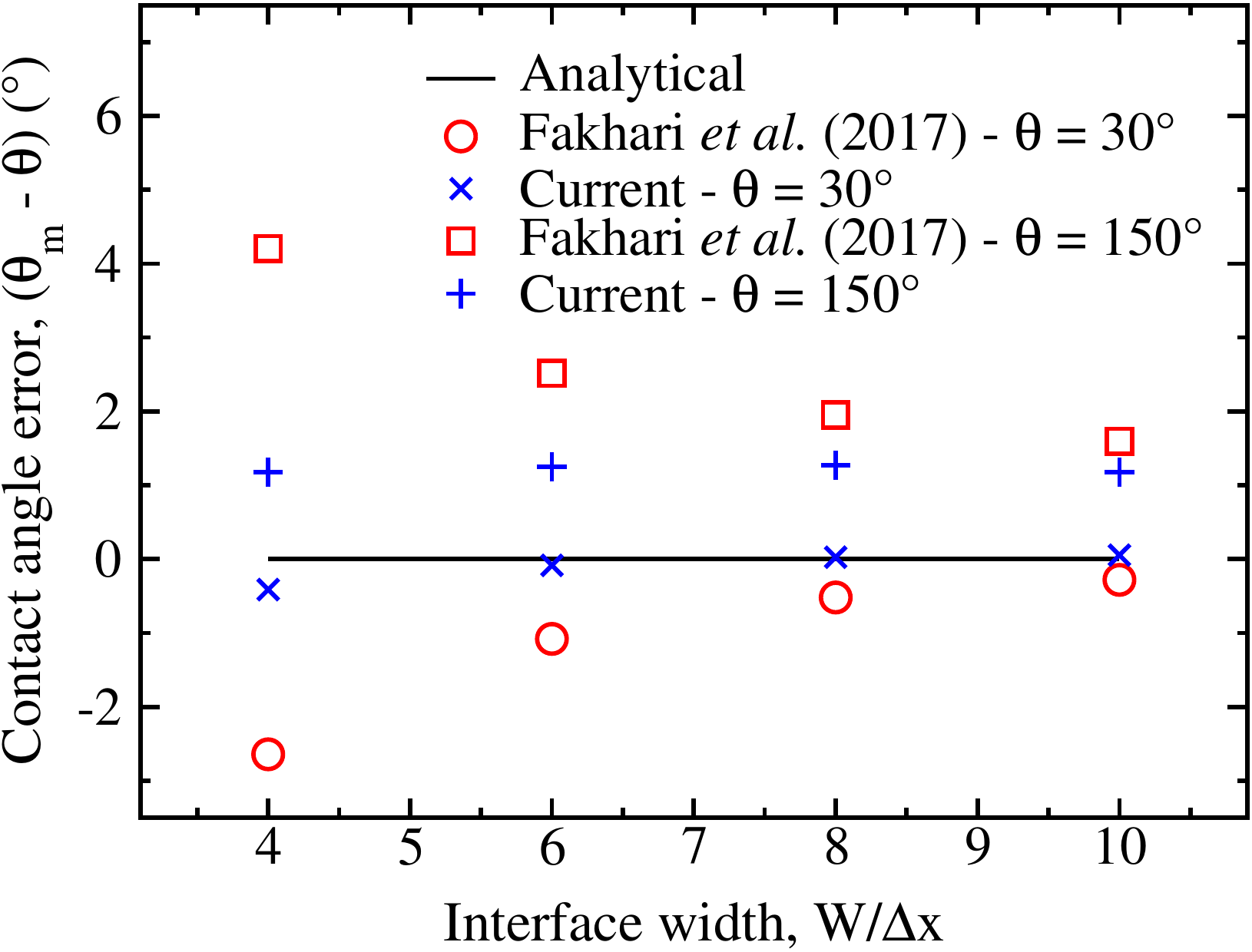}
    \caption{Error in contact angle as a function of interface width $W$. For a given contact angle $\theta$, increasing $\frac{W}{\Delta x}$ effectively decreases magnitude of the small parameter $|\varepsilon| = \frac{4\Delta x}{W}|\cos\theta|$. For a moderate grid resolution $\frac{W}{\Delta x}=4$, the present wetting scheme performs better than the scheme of Fakhari \textit{et al.}~\cite{fakhari2017diffuse}.  }
    \label{fig:dropPlateErrorScalingWidth}
\end{figure}

Interface width $W$ is an auxiliary parameter in the model. Cahn number $\text{Cn} = \frac{W}{L}$  dictates applicability of the diffuse interface framework, where $L$ is the smallest characteristic length scale of the system. Smaller values of $W$ cause higher discrete errors in the evaluation of the order parameter gradient, while higher $W$ requires a larger computational domain to maintain the same Cahn number. For this benchmark, a common characteristic length scale $L$ is defined as $L=\sqrt{A/\pi}$. We vary the interface width from $W=4$ to $W=10$, which corresponds to a Cahn number variation of {$0.07\le\text{Cn}\le 0.17$} for contact angles $\theta=30\degree$ and $\theta=150\degree$ and plot the associated error in Fig.~(\ref{fig:dropPlateErrorScalingWidth}). Both the current and Fakhari \textit{et al.}~\cite{fakhari2017diffuse} schemes produce less than $2\degree$ errors when $W\ge6$. For the remaining benchmarks, interface width is kept constant as $W=4$.

\subsection{Drop on solid cylinder}
\label{subsection:dropCylinder}
\begin{figure}[htbp]
    \centering
    \includegraphics[width=0.8\linewidth]{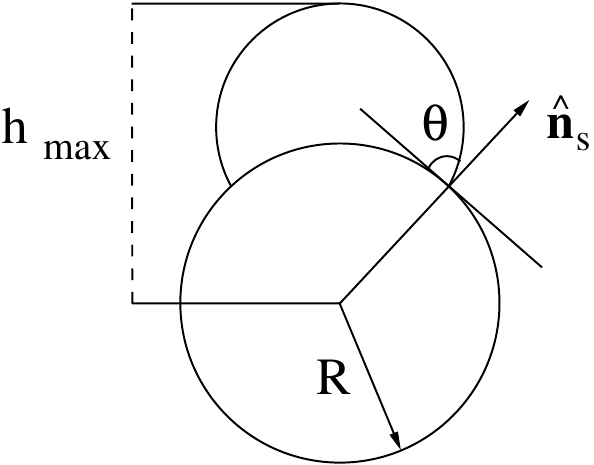}
    \caption{Schematic of a stationary drop sitting on a solid cylinder of radius $R$ under negligible gravity. The contact angle is $\theta$, and the height of the apex of the droplet at steady state is $h_\text{max}$.}
    \label{fig:dropCylinderSchematic}
\end{figure}

Consider a stationary drop sitting on a solid cylinder of radius $R$ under negligible gravity as shown in Fig.~(\ref{fig:dropCylinderSchematic}). For a given equilibrium contact angle $\theta$ and drop area $A$, and assuming the drop remains circular, the height of the apex of the droplet measured from the center of the solid cylinder $h_\text{max}$ can be found analytically~\cite{xu2017lattice,fakhari2017diffuse,o2018volume}. In this section, we simulate a drop sitting on a cylinder for various contact angles, and compare $h_\text{max}$ measured from simulations with the analytical solution.  

We consider a square computational domain $(L_x,  L_y) = (5 R, 5 R)$ where $R = 40$, with the cylinder center fixed at $(L_x/2, L_y/4)$. To reduce computational time, we initialize the drop with the expected equilibrium configuration for a given area $A = 3000$. All boundaries of the computational domain are modeled as periodic. We perform simulations for various contact angles in the range $30\degree-150\degree$, and compare $h_\text{max}$ measured from simulations with the analytical solution. 

\begin{figure}[htbp]
    \centering
    \includegraphics[width=\linewidth]{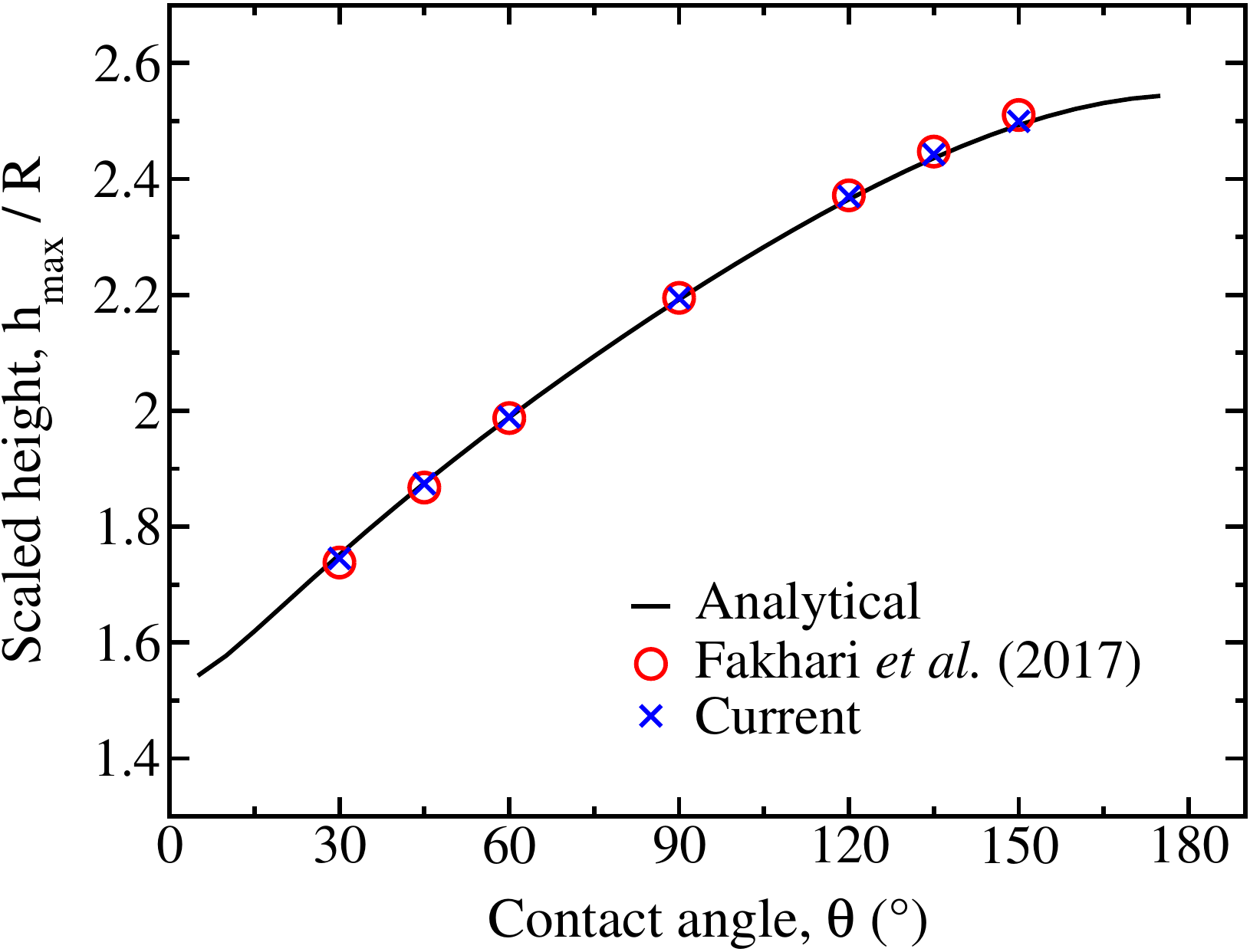}
    \caption{Scaled height of the drop $\frac{h_{\text{max}}}{R}$ along $y$ axis as a function of contact angle $\theta$. As the contact angle $\theta$ deviates from the neutral contact angle, the present wetting scheme produces better accuracy, albeit slightly.}
    \label{fig:dropCylinderHeightVsTheta}
\end{figure}

Figure~(\ref{fig:dropCylinderHeightVsTheta}) compares the height of the drop measured from the cylinder center obtained using the proposed wetting scheme against Fakhari~\textit{et al.}'s~\cite{fakhari2017diffuse} scheme, as a function of contact angle. Both the schemes produce close agreement with the analytical solution for intermediate contact angles $(60\degree-120\degree)$. For $\theta = 30\degree$ and $\theta = 150\degree$, Fakhari~\textit{et al.}'s~\cite{fakhari2017diffuse} scheme makes $-0.76\%$ and $0.7\%$ errors in scaled height $h_\text{max}/R$ respectively. For the same contact angles, the proposed scheme produces $-0.35\%$ and $0.26\%$ errors, respectively.
% Some deviations can be observed at extreme contact angles $\theta=30\degree$ or $\theta=150\degree$ for Fakhari~\textit{et al.}'s~\cite{fakhari2017diffuse} scheme, whereas the proposed scheme stays accurate. \textcolor{red}{why curved geometries makes two schemes equivalent? flat boudary clearly shows improvement of current scheme over Fakhari et al}

\subsection{Fixed particle at liquid-gas interface}
\label{subsection:particleInterface}
\begin{figure}[htbp]
    \centering
    \includegraphics[width=\linewidth]{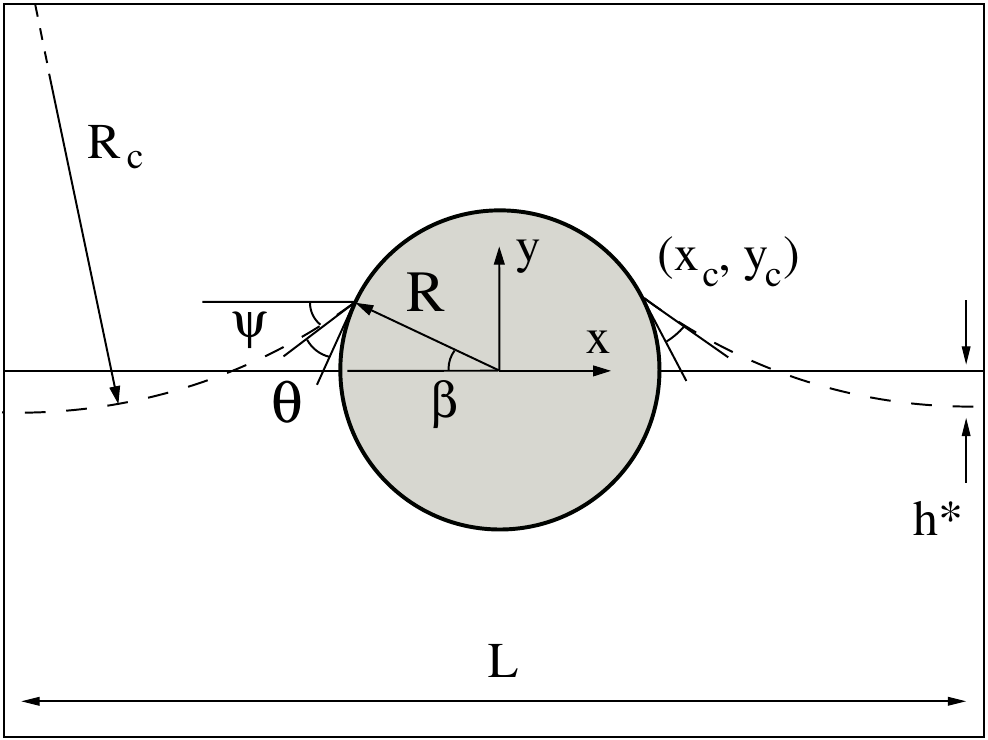}
    \caption{A cylindrical particle of radius $R$ is placed at a flat liquid-gas interface such that center of the particle is on the interface. Periodic and no-slip boundary conditions are applied in the horizontal $x$ and vertical $y$ directions, respectively.  Depending on the contact angle $\theta$, the contact line will adjust its position on the particle while conserving mass of both the fluid phases.}
    \label{fig:particleInterfaceSchematic}
\end{figure}

Particle arrangements at fluid-fluid interfaces are common in multi-phase suspensions~\cite{guzman2022broad}. In these systems, wettability controls both the capillary forces acting on the particles and the equilibrium shape of the fluid-fluid interface. To demonstrate the efficacy of the proposed wetting scheme for such problems, we simulate a periodic array of fixed solid cylinders of radius $R$ suspended at a liquid-gas interface under negligible gravity, as shown in Fig.~(\ref{fig:particleInterfaceSchematic}).

The periodic domain has length $L$, with a cylinder centered at $(L/2, L/2)$. Initially, the liquid-gas interface is flat and passes through the cylinder center, i.e., $y=L/2$. When the cylinder is held fixed, the contact line moves along the cylinder until the prescribed equilibrium contact angle $\theta$ is reached. For $\theta<90\degree$, the interface climbs upward on the cylinder, while for $\theta>90\degree$, it moves downward. Given the boundary conditions, respective amounts of liquid and gas in the system are conserved. Therefore, for $\theta \neq 90\degree$, the equilibrium interface deforms to satisfy both the contact-angle condition and mass conservation. The setup is equivalent to an infinite periodic array of cylinders trapped at a liquid-gas interface, where particles are held fixed, for example, with optical tweezers~\cite{kotnala2017microfluidic}. 

We describe the equilibrium geometry for the case $\theta<90\degree$. The contact-line position on the cylinder is specified by the half-filling angle $\beta$. Further, the liquid-gas interface makes an angle $\psi$ with the horizontal at the contact line. The deformed interface is assumed to be a circular arc with radius of curvature $R_\text{c}$. Due to symmetry, it is sufficient to consider the half-domain $x\le L/2$. The center of curvature of the interface lies on the periodic boundary, at either $(L, R\sin\psi + R\sin\beta)$ or equivalently $(0, R\sin\psi + R\sin\beta)$. Using this geometric constraint together with mass (area) conservation of fluid phases, the unknowns $R_\text{c}$, $\psi$, and $\beta$ are determined from
\begin{align}
    & R^2 \Big ( \frac{L}{R} \sin{\beta} - \sin{\beta} \cos{\beta} -\beta \Big ) = R_\text{c}^2(\psi - \sin{\psi} \cos{\psi}), \nonumber \\
    & 2R_\text{c}\cos{(\beta + \theta)} = {L - 2 R \cos{\beta}}, \nonumber \\
    & \psi = \frac{\pi}{2} - (\beta + \theta), \label{eq:dropAtInterfaceConstraints}
\end{align}
solving which completely determines the steady-state interface shape, for given $L$, $R$ and $\theta$. In this section, we compare the vertical displacement $h^*$ of the liquid-gas interface from the initial position at the periodic boundary $x = 0$ or equivalently $x = L$, obtained using both the wetting schemes for various contact angles $\theta$ with the analytical solution given as $h^* = R_\text{c} (1 - \cos{\psi}) - R \sin{\beta}$.

\begin{figure}[htbp]
    \centering
    \includegraphics[width=\linewidth]{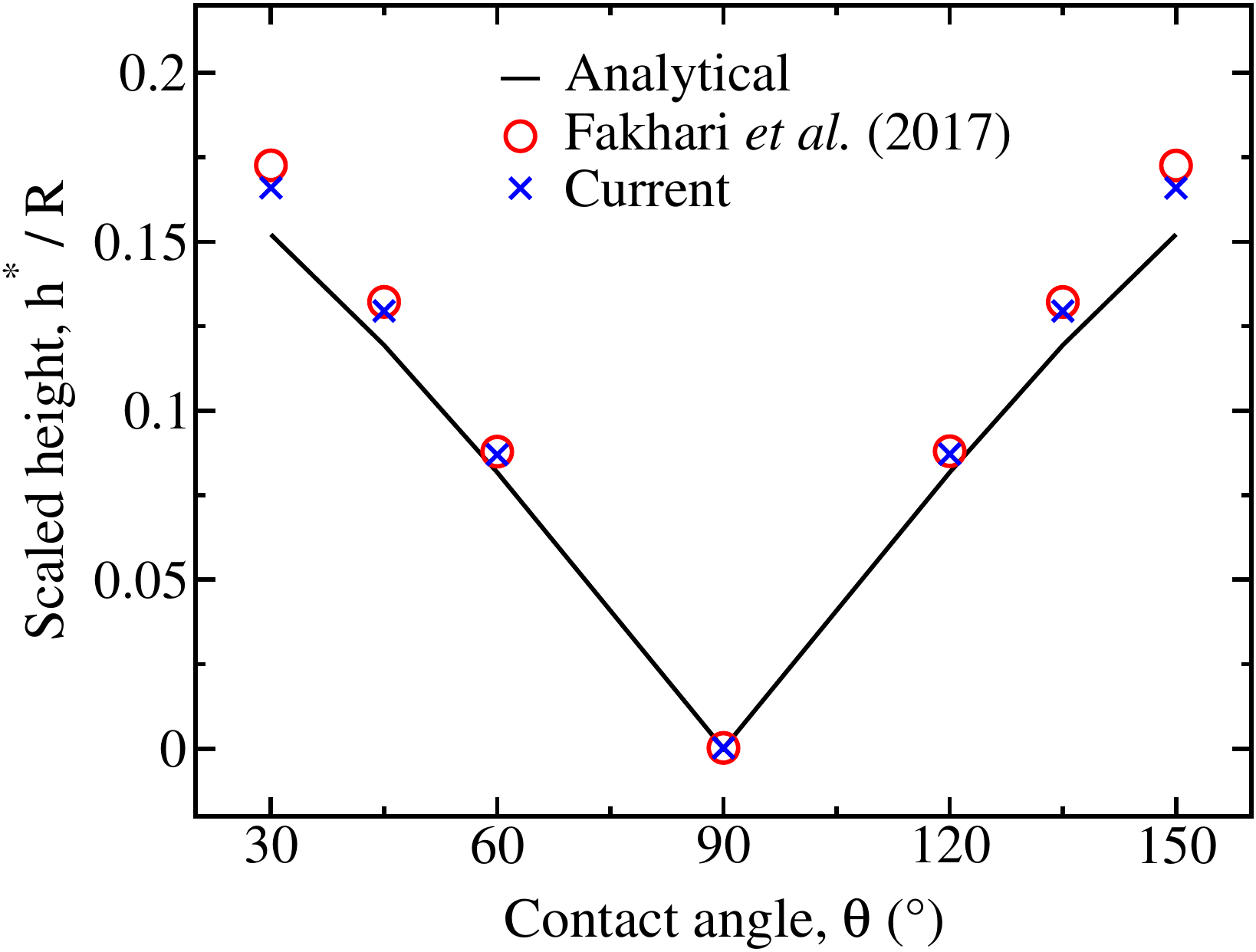}
    \caption{Vertical displacement $h^*$ of the liquid-gas interface from the initial position at the periodic boundary $x = 0$ as a function of contact angle $\theta$. $h^*$ is obtained by numerical solution of Eq.~(\ref{eq:dropAtInterfaceConstraints}).}
    \label{fig:particleInterface_h_vs_theta}
\end{figure}

The computational domain is a square box of size $(L_x, L_y) = (4.8 R,4.8 R)$ with $R = 25$, and cylinder centered at $(L_x/2, L_y/2)$. The domain is initially filled with liquid up to $L_y/2$ from the bottom boundary. The top and bottom boundaries of the domain are modeled as no-slip walls, while the left and right boundaries are periodic.  We simulate the system until a steady state is reached for a given equilibrium contact angle $\theta$, after which $h^*$ is obtained from interpolation.

Figure~(\ref{fig:particleInterface_h_vs_theta}) compares the vertical displacement $h^*$ of the liquid-gas interface from the initial position at the periodic boundary $x = 0$ as a function of $\theta$, as obtained from the proposed wetting scheme against Fakhari~\textit{et al.}'s~\cite{fakhari2017diffuse} scheme. At extreme contact angles $30\degree$ and $150\degree$, the error in vertical displacement from the proposed scheme is $\sim 9\%$, compared to $\sim 14\%$ from Fakhari~\textit{et al.}'s~\cite{fakhari2017diffuse} scheme. For moderate contact angles $60\degree-120\degree$, the proposed wetting scheme produces $\sim 6\%$, compared to $\sim 8\%$ from Fakhari~\textit{et al.}'s~\cite{fakhari2017diffuse} scheme. Thus, for all contact angles considered, the proposed wetting scheme produces lesser error compared to Fakhari~\textit{et al.}'s~\cite{fakhari2017diffuse} scheme.

\subsection{Drop impact on cylinder}
\label{subsection:dropImpact}

\begin{figure}[htbp]
    \centering
    \includegraphics[width=0.8\linewidth]{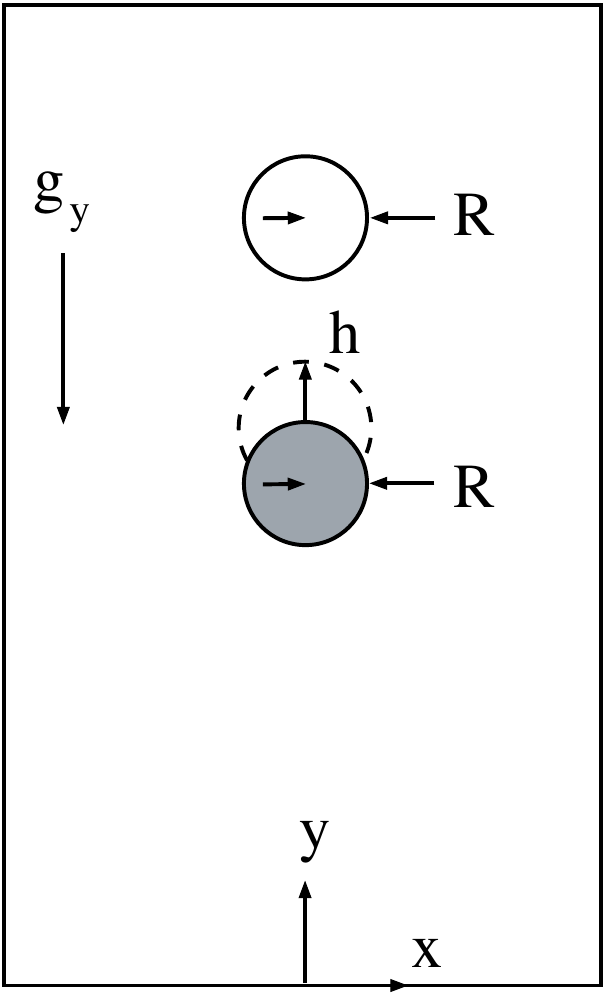}
    \caption{Schematic of a drop of initial radius $R_{\text{d}}$ falling on a fixed cylinder of radius $R_{\text{s}}$, due to gravity. The height of the evolving film $h$ is obtained from simulations, and compared against analytical scaling law~\cite{yarin1995impact, bakshi2007investigations}.}
    \label{fig:dropImpact_schematic}
\end{figure}

\begin{figure*}[htbp]
    \centering
    \includegraphics[width=\linewidth]{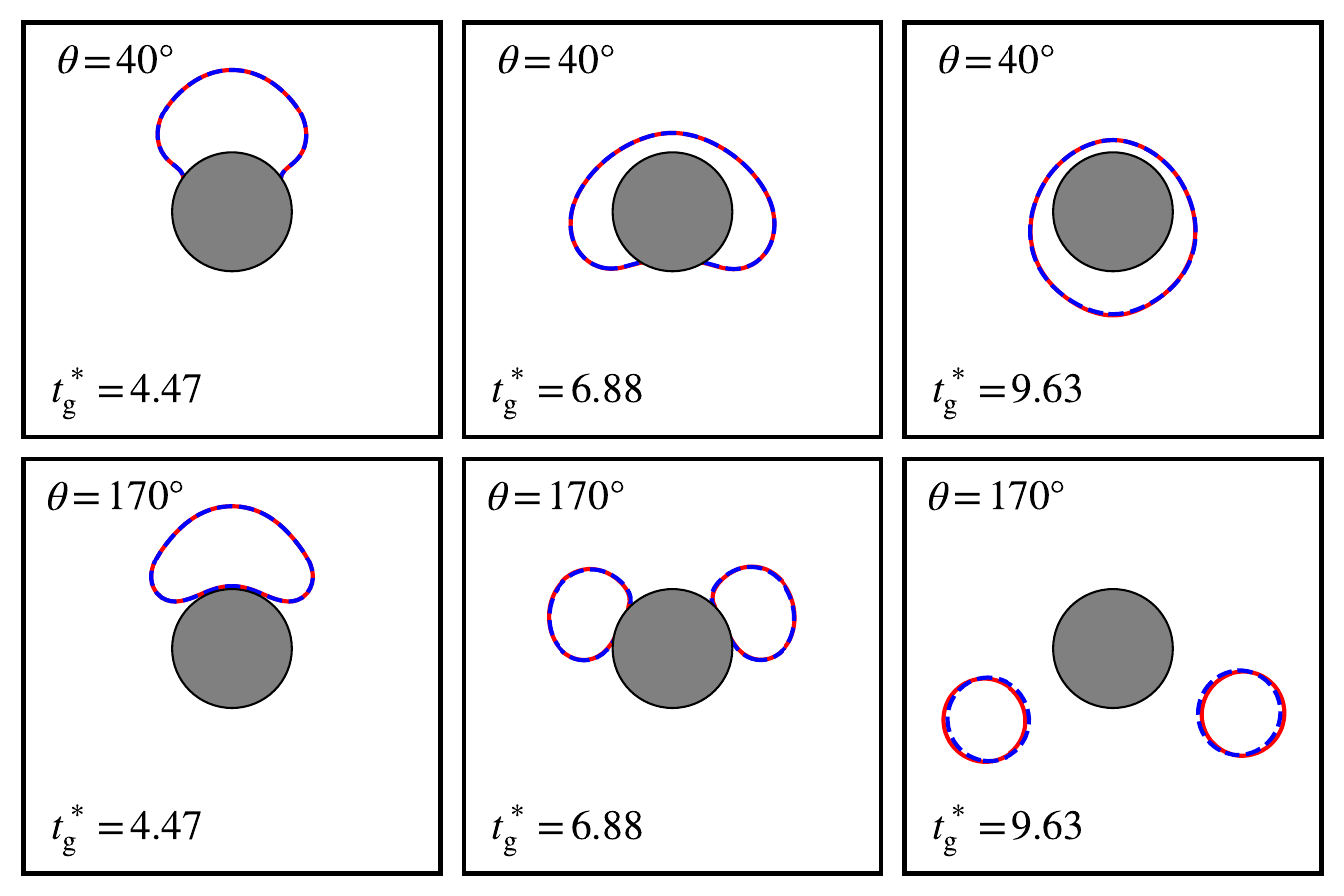}
    \caption{Contours of $\varphi = 0.5$ showing the drop shapes obtained using the proposed scheme (blue dashed line) and Fakhari \textit{et al.}'s~\cite{fakhari2017diffuse} (solid red line) scheme for Bo = 2.2, and $\theta = 40\degree$ and $170\degree$ at various time-instants. $t^*_\text{g} = t \sqrt{g / 2 R}$ is time scaled with gravitational time scale.}
    \label{fig:dropImpact_contours}
\end{figure*}

In the previous benchmarks, we compared simulated contact angles and liquid-gas interface locations with theory in the steady state. Here, we compare the dynamically evolving liquid-gas interface position while in contact with a stationary solid at high Reynolds and Bond numbers with other numerical studies. Specifically, we simulate drop impact on a cylinder~\cite{o2018volume,yao2022multiphase}. Figure~(\ref{fig:dropImpact_schematic}) shows a schematic of the problem.

Consider a rectangular domain of size $(L_x, L_y) = (L,2L)$ and particle size length scale $R=L/10$. Here, $L_{x}$ and $L_{y}$ are the dimensions of the system in the $x$ and $y$ directions, respectively.
A fixed solid cylinder of radius $R_{\text{s}}=R=L/10$ is placed at ($L_x/2, L_{y}/2$), in an otherwise quiescent gas.   A liquid drop of radius $R_{\text{d}} = R = L/10$, with initial position of $(L_x/2, 3L_{y}/4)$, starts to move downwards towards the fixed cylinder under gravity $\mathbf{g} = (0,-g_y)$. The fluid density and viscosity ratios are $\frac{\rho_{\text{l}}}{\rho_{\text{g}}} = 1000$ and $\frac{\mu_{\text{l}}}{\mu_{\text{g}}} = 100$, respectively. The density and dynamic viscosity of the gas phase are $\rho_\text{g} = 5.997$ and $\mu_\text{g} = 2.326 \times 10^{-2}$ respectively, which are different than the default values. The dimensionless numbers characterizing the system are Reynolds number Re $ = \rho_\text{l} U (2 R) / \mu_\text{l} =25$ and Bond number Bo $= g_y (\rho_\text{l} - \rho_\text{g}) (2 R)^2 / \sigma = 2.2$, where $U = \sqrt{2 g_y R}$. Given the low capillary number Ca $= \mu_\text{l} U / \sigma = 0.088$, the drop retains approximately circular shape while gaining kinetic energy as it descends towards the solid cylinder. As the drop contacts the fixed solid cylinder, the contact angle $\theta$ controls the shape of the drop. We consider a hydrophilic cylinder with $\theta=40\degree$ and a hydrophobic cylinder with $\theta=170\degree$. For the hydrophilic cylinder, the liquid-gas interface spreads along the curved solid boundary after impact. The drop remains attached to the cylinder and engulfs the entire solid surface.

\begin{figure}[htbp]
    \centering
    \includegraphics[width=1\linewidth]{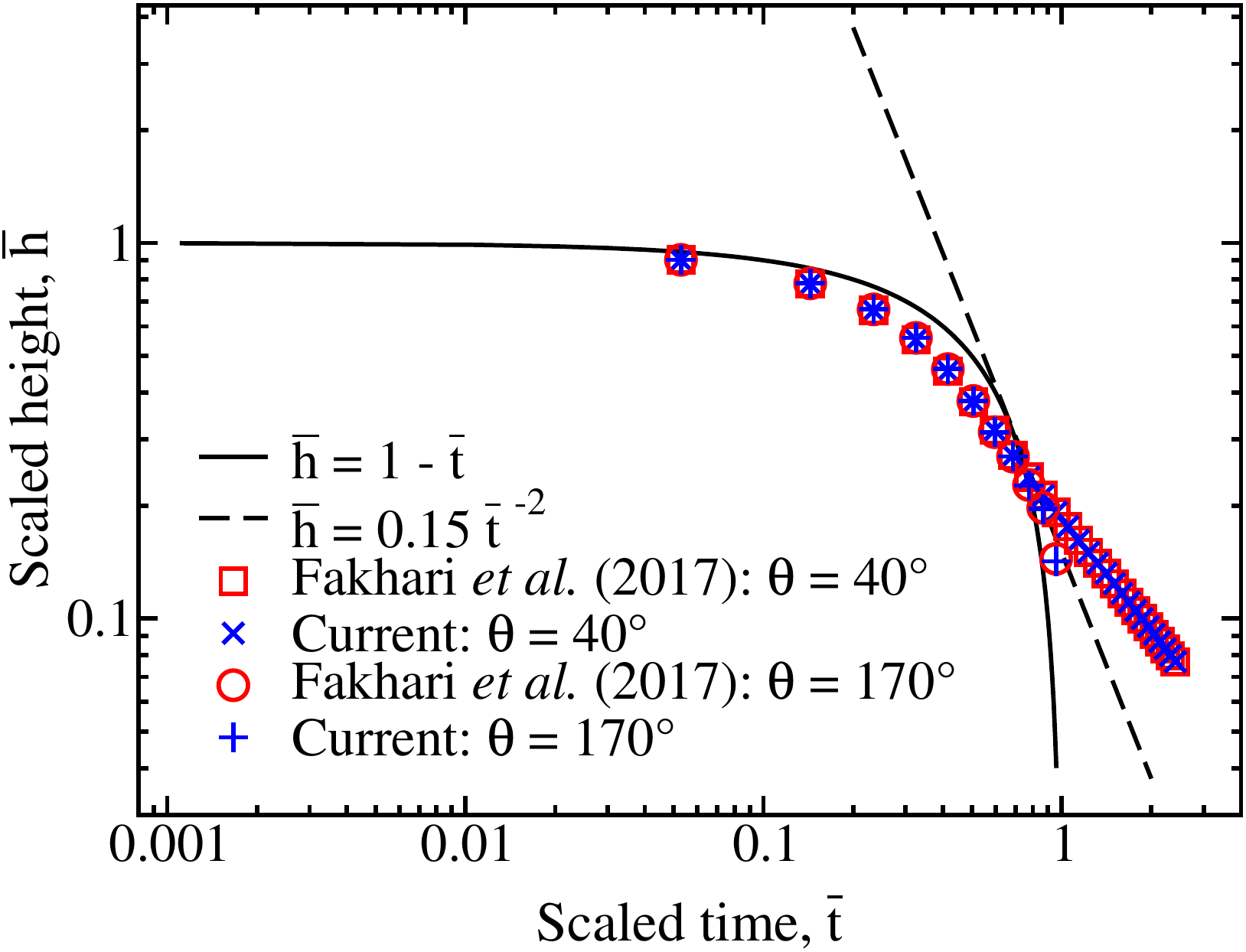}
    \caption{Scaled height $\overline{h}$ of the drop film on the cylinder surface as a function of scaled time $\overline{t}$. Here, $\overline{h} = h / 2 R$, where $h$ is the measured film height in the simulations, and $\overline{t} = (t - t_\text{i}) U_\text{i} / 2 R$, where $t_\text{i}$ is the time of impact and $U_\text{i}$ is the impact velocity. Both, $U_\text{i}$ and $t_\text{i}$, are estimated by relating the initial potential energy of the drop with the final kinetic energy at the instant of impact.}
    \label{fig:dropImpact_h_vs_t}
\end{figure}

For the contact angle $\theta=170\degree$, the drop shape differs after contact with the solid cylinder. Energetically, the drop tries to minimize contact area with the solid cylinder. Coupled with the inertia of the drop, this results in the drop splitting in two. For both the contact angle cases, drop shapes obtained with the scheme of Fakhari~\textit{et al.}~\cite{fakhari2017diffuse} and the current wetting scheme are compared in Fig.~(\ref{fig:dropImpact_contours}). On the scale of the drop, the two schemes lie on top of each other. For a quantitative comparison, we compare the film thickness of the drop along the centerline of the solid cylinder in Fig.~(\ref{fig:dropImpact_h_vs_t}). Both the hydrophilic case with $\theta=40\degree$ and the hydrophobic case with $\theta=170\degree$ follow the expected analytical scaling~\cite{yarin1995impact,bakshi2007investigations} for early and intermediate-time film thickness. However, the agreement with experiments is qualitative as the simulations are two-dimensional and performed at lower Reynolds and Weber numbers than three-dimensional experiments on a sphere. The two wetting schemes are nearly identical in film thickness as a function of time.

\subsection{GPU realization}
\label{section:GPU}

Use of JAX~\cite{jax2018github} for high-performance numerical simulations in fluid mechanics has been reported~\cite{ataei2024xlb,BEZGIN2025109433}. We use JAX as the computational backend, which combines automatic differentiation with just-in-time compilation through the XLA compiler~\cite{openxla2026}. Such construction allows high-level tensor operations to be lowered to optimized GPU kernels. Kernel fusion enables multiple computational steps to be executed as a single optimized operation, leading to improved memory efficiency.  

The solver is formulated within a functional program, separating constant and evolving quantities into \textit{System} and \textit{State}, respectively. The \textit{System} includes constant parameters such as densities of immiscible phases, surface tension, reference pressure and lattice of choice, while the \textit{State} contains dynamic fields including fluid velocity, pressure, phase field, and distribution functions. All lattice Boltzmann operations are implemented as pure, grid-local functions and vectorized across the computational domain. Each iteration consists of collision and streaming steps expressed as tensor operations and compiled into optimized kernels. 

Performance is evaluated on an NVIDIA A100 GPU for the benchmarks described in the previous section. Efficiency is reported in terms of million lattice updates per second (MLUPS), defined as
 \begin{align}
\text{MLUPS} = \frac{\text{grid size}\times \text{number of iterations}}{\text{execution time}\times 10^6}. \label{eq:MLUPSDefinition}
\end{align}

We report MLUPS values, as defined in Eq.~(\ref{eq:MLUPSDefinition}), averaged over 10 runs.
The compilation overhead associated with JIT is excluded by omitting the first iteration. The results show that small grid sizes underutilize GPU resources, while increasing the grid size improves performance until saturation is reached. On average, single precision FP32 computations are at least two times faster than the double precision FP64 ones as shown in Table~\ref{tab:performance-all}. For the first three benchmarks, FP32 and FP64 produced same results. However, FP32 computations become unstable for the drop-impact-on-cylinder benchmark~\ref{subsection:dropImpact}. We attribute this behavior to relatively high Reynolds and Weber number of the problem. 

\begin{table}[!htp]
  \caption{MLUPS comparison between FP32 and FP64 for $50,000$ iterations, averaged over 10 runs}
  \label{tab:performance-all}
  \begin{tabular}{llll}
    \toprule
    {Benchmark} &  {\shortstack{Grid  Size}} & {\shortstack{FP32 }} & {\shortstack{FP64 }} \\
    \midrule
    \multirow{3}{*}{\shortstack{Drop on  \\ flat plate}}
    &   $500\times300$   & 759  & 285  \\
    &   $1000\times600$  & 877  & 257  \\
    &  $1500\times900$ & 768  & 263  \\
    \midrule
    \multirow{3}{*}{\shortstack{Drop on  \\ cylinder}}
    &  $201\times201$  & 374  & 216  \\
    &  $401\times401$  & 740  & 310  \\
    &  $601\times601$ & 844  & 283  \\
    \midrule
    \multirow{3}{*}{\shortstack{Particle at liquid-\\ gas interface}}
    &  $120\times122$ & 175  & 110  \\
    &  $240\times242$ & 466  & 225  \\
    &  $360\times362$ & 748  & 287  \\
    \midrule
    \multirow{3}{*}{\shortstack{Drop impact\\ on cylinder}}
    &  $128\times258$ & --  & 191  \\
    &  $256\times514$ & --  & 293  \\
    &  $384\times770$ & --  & 289  \\
   \bottomrule
\end{tabular}
\end{table}

\section{Summary and Conclusions}\label{section:conclusion}
We present a wetting boundary condition for diffuse liquid-gas interfaces. The wetting scheme updates the solid boundary (ghost) nodes inside the solid phase. The ghost node update rule extends the equilibrium order parameter (color-field) profile into the solid region. The scheme is verified against analytical solutions and other numerical schemes with various benchmarks.

Numerical results demonstrate that the wetting scheme accurately handles a wide range of contact angles on curved solid boundaries in both static and dynamic situations. Diffuse interface models introduce surface tension effect via a distributed force on the liquid-gas interface, leading to spurious (parasitic) fluid velocity currents in the interfacial region. In the presence of solid boundaries, these spurious currents usually increase, especially in the vicinity of the contact line. Typical magnitude of spurious currents with the present wetting boundary scheme remains of the order of $\mathcal{O}(10^{-6})$ in lattice units that are a couple of orders of magnitude lower than the popular Pseudopotential lattice Boltzmann models~\cite{xu2025mitigating}. The wetting prescription depends on the interface grid resolution $\varepsilon$. A series expansion in $\varepsilon$ shows that the present prescription approaches Fakhari \textit{et al.}’s prescription~\cite{fakhari2017diffuse} in the high grid resolution limit ($\varepsilon\to0$), while avoiding the special treatment required for the neutral contact angle $\theta=90\degree$. GPU simulations also confirm that the use of single-precision format FP32 offers a significant increase in computational efficiency over double-precision FP64 format. For test cases used in this study, both formats produce nearly identical results, except when the single-precision format FP32 simulations produced undefined numbers for high Reynolds and Weber number drop impact test~\ref{subsection:dropImpact}. 

Although the scheme is applied here to a color-gradient model, it can be used directly for phase-field~\cite{fakhari2017improved} or free-energy~\cite{lee2008wall} models. The wetting scheme presented here can be directly extended to three dimensions~\cite{younes2022phase}.  The present numerical framework is currently restricted to single-GPU execution. Extension of the present numerical framework to Multi-GPU parallelization, along with inclusion of checkpointing feature using the Orbax library~\cite{orbax2024}, will be addressed elsewhere.  Coupled with recent advances in the computation of contact line capillary force~\cite{bhattacharya2026Capillary} on solid particles and in hysteresis schemes for mobile solid particles~\cite{bhattacharya2026hysteresis}, the proposed method provides a robust framework for simulating systems involving solid particles at liquid–gas interfaces.

\begin{acknowledgments}
M.~B. and A.~S. acknowledge support by the Indian Institute of Technology  Bombay (IITB) through Seed Grant Fund: RD/0522-IRCCSH0-008. Simulations were performed on Praganak and Prajna High Performance Computing facility at IITB.
\end{acknowledgments}

\appendix

\section{Lattice Boltzmann model}
The color-gradient model is described here in brief. The blue $B_i$ and red $R_i$ colored fluid populations represent the liquid and gas phases, respectively. Collision rule for the color blind fluid populations $N_i = B_i + R_i$ is given as   \begin{equation}
 \begin{split}
  N^*_i &= N_i -(\mathbf{M}^{-1}\mathbf{S}\mathbf{M})_{ij}(N_j- N^{\text{eq}}_j) \\
  &+ \left[\mathbf{M}^{-1}\left(\mathbf{I}-\frac{\mathbf{S}}{2}\right)\mathbf{M}\right]_{ij} F_j, \label{Eq:MRTCOLL}
 \end{split}
\end{equation}
where the transformation matrix $\mathbf{M}$ and  the diagonal relaxation matrix $\mathbf{S}$, respectively are ~\cite{subhedar2022color}
\begin{align}
\renewcommand{\arraystretch}{1.2}
\mathbf{M}=\left(
\begin{array}{ccccccccc}
 1 & 1 & 1 & 1 & 1 & 1 & 1 & 1 & 1 \\
 -4 & -1 & -1 & -1 & -1 & 2 & 2 & 2 & 2 \\
 4 & -2 & -2 & -2 & -2 & 1 & 1 & 1 & 1 \\
 0 & 1 & 0 & -1 & 0 & 1 & -1 & -1 & 1 \\
 0 & -2 & 0 & 2 & 0 & 1 & -1 & -1 & 1 \\
 0 & 0 & 1 & 0 & -1 & 1 & 1 & -1 & -1 \\
 0 & 0 & -2 & 0 & 2 & 1 & 1 & -1 & -1 \\
 0 & 1 & -1 & 1 & -1 & 0 & 0 & 0 & 0 \\
 0 & 0 & 0 & 0 & 0 & 1 & -1 & 1 & -1 \\
\end{array}
\right),
\end{align}
\begin{equation}
\mathbf{S} = \text{diag}(1,1,1,1,1,1,1,s_7,s_8),
\end{equation}
where $s_7 = s_8 = \frac{1}{\tau}$ and $\tau = 0.5 + \frac{\nu}{\Delta tc^2_{\text{s}}}$ is relaxation time, where $\Delta t = 1$ is unit time-step. The color-blind equilibrium distribution function is 
\begin{equation}
N^{\text{eq}}_i  = 
\begin{cases}
 \chi_i-(1-w_i)\frac{p}{c^2_{\text{s}}} -w_i\frac{\mathbf{u}^2}{2c^2_{\text{s}}} \hfill i=0\\
 \chi_i +w_i \left[  \frac{p}{c^2_{\text{s}}} + \frac{\mathbf{c}_i\cdot\mathbf{u}}{c^2_{\text{s}}} + \frac{(\mathbf{c}_i\cdot\mathbf{u})^2}{2c^4_{\text{s}}}-\frac{\mathbf{u}^2}{2c^2_{\text{s}}}\right] \hfill i\ne0,
\end{cases}\label{Eq:equilibriumDistribution}
\end{equation}
where $w_i$ is lattice weight, $\mathbf{c}_i$ is lattice velocity, $c_{\text{s}}$ is lattice speed of sound, $\mathbf{u}$ is fluid velocity and $p=\frac{p_{\text{h}}}{\rho}$ is auxiliary pressure field.  For the D2Q9 stencil chosen here, the lattice speed of sound $c_{\text{s}}=\frac{1}{\sqrt{3}}$. The lattice velocities $\mathbf{c}$, lattice weights $w$ and color weights $\chi$ are as follows

\begin{align}
\mathbf{c}=
\left(
\begin{array}{ccccccccc}
 0 & 1 & 0 & -1 & 0 & 1 & -1 & -1 & 1 \\
 0 & 0 & 1 & 0 & -1 & 1 & 1 & -1 & -1 \\
\end{array}
\right), 
\end{align}

\begin{equation}
w_i = 
\begin{cases}
			\frac{4}{9} & {i=0}\\
            \frac{1}{9} & {i=1,2,3,4}\\
            \frac{1}{36}&  {i=5,6,7,8}\
		 \end{cases},
\end{equation}
\begin{equation}
\chi_i = 
\begin{cases}
			\alpha& {i=0}\\
            \frac{1-\alpha}{5} & {i=1,2,3,4}\\
            \frac{1-\alpha}{20}&  {i=5,6,7,8}
		 \end{cases},\label{Eq:ChiTerm}
\end{equation}
where $\alpha$ controls liquid-gas interface mobility $M_{\varphi} =  \frac{c^2_{\text{s}} \Delta t}{2}\left[\frac{9(1-\alpha)}{5}+\frac{p}{c^2_{\text{s}}}\right] $.
The force term  $F_i$ in Eq.~(\ref{Eq:MRTCOLL}) is given by~\cite{zu2013phase}
\begin{equation}
F_i = \frac{w_i }{c^2_{\text{s}}} \frac{\mathbf{F}_{\text{macro}} \Delta t}{\rho}\cdot \mathbf{c}_i, \label{Eq:forceGuo}
\end{equation}
where $\mathbf{F}_{\text{macro}}$  is the total force acting on the fluid that contains both the correction and physical forces~\cite{zu2013phase}
\begin{equation}
\mathbf{F}_{\text{macro}} = \mathbf{F}_{\text{visc}} + \mathbf{F}_{\text{p}} + \mathbf{F}_{\text{s}} + \mathbf{F}_{\text{b}}. \label{Eq:totalForce}
\end{equation}
 $\mathbf{F}_{\text{b}}$  and $\mathbf{F}_{\text{s}}$ in Eq.~(\ref{Eq:totalForce}) represent the  body and surface tension forces. The surface tension force $\mathbf{F}_{\text{s}}$ is given by Eq.~(\ref{eq:surfaceTensionForce}). In addition, $\mathbf{F}_{\text{visc}}$  and $\mathbf{F}_{\text{p}}$ are the viscous correction force and  is pressure correction force, respectively. The expressions for these forces are as follows~\cite{zu2013phase}
\begin{align}
\mathbf{F}_{\text{p}}  &= -p{\nabla \rho}, \label{Eq:pressCorrection}\\
\mathbf{F}_{\text{visc}} & = c^2_{\text{s}}\Delta t\left(\uptau-\frac{1}{2} \right)({\nabla \mathbf{u} + \nabla \mathbf{u}^{\text{T}}})\cdot \nabla \rho. \label{Eq:viscCorrection}
\end{align}

The segregation rule for the liquid phase is
\begin{align}
B^{*}_{i} &= \varphi N^*_i + 2\frac{\mathbf{c}_i\cdot\mathbf{\hat n}_{\varphi}}{W}\varphi(1-\varphi) N^{\text{eq},\mathbf{u=0}}_i, \label{Eq:segragationStepBlue}
\end{align}
where $B^{*}_{i} $ is the post-segregation population of the blue (liquid) phase. Further, $N^{\text{eq},\mathbf{u=0}}_i $ is the equilibrium distribution function evaluated for $\mathbf{u} = 0$. The color, fluid velocity and auxiliary pressure field are computed as
\begin{align}
\varphi &= \sum_i B_i, \label{Eq:phiUpdate}\\
\mathbf{u} &= \sum_i N_i\mathbf{c}_i + \frac{\mathbf{F}_{\text{macro}}\Delta t}{2\rho},\label{Eq:velUpdate}\\
p &= \frac{1}{1-w_0} \left(-\frac{w_0}{2}\mathbf{u}^2 - (1-\chi_0)c^2_{\text{s}}+ \sum_{i\ne0}N_ic^2_{\text{s}}\right). \label{Eq:pressureUpdate}
\end{align}

It is enough to execute the propagation step for color-blind and liquid phase fluid populations  as

\begin{align}
N_i(\mathbf{x}+\mathbf{c}_i\Delta t, t+\Delta t) = N_i^*(\mathbf{x},t), \label{Eq:propTotal}\\
B_{i}(\mathbf{x}+\mathbf{c}_i\Delta t, t+\Delta t) = B_{i}^{*}(\mathbf{x},t).\label{Eq:propBlue}
\end{align}

% \nocite{TitlesOn}
% \bibliographystyle{apsrev4-2}
\bibliography{ref}

@article{utada2007dripping,
  title={Dripping, jetting, drops, and wetting: The magic of microfluidics},
  author={Utada, AS and Chu, L-Y and Fernandez-Nieves, A and Link, DR and Holtze, C and Weitz, DA},
  journal={Mrs Bulletin},
  volume={32},
  number={9},
  pages={702--708},
  year={2007},
  publisher={Cambridge University Press},
  doi={10.1557/mrs2007.145}
}

@article{BEZGIN2025109433,
title = {JAX-Fluids 2.0: Towards HPC for differentiable CFD of compressible two-phase flows},
journal = {Computer Physics Communications},
volume = {308},
pages = {109433},
year = {2025},
issn = {0010-4655},
doi = {10.1016/j.cpc.2024.109433},
author = {Deniz A. Bezgin and Aaron B. Buhendwa and Nikolaus A. Adams},

}

@article{ataei2024xlb,
title = {XLB: A differentiable massively parallel lattice Boltzmann library in Python},
journal = {Computer Physics Communications},
volume = {300},
pages = {109187},
year = {2024},
issn = {0010-4655},
doi = {10.1016/j.cpc.2024.109187},
author = {Mohammadmehdi Ataei and Hesam Salehipour},

}

@misc{openxla2026,
  title        = {{XLA}: Accelerated Linear Algebra},
  author       = {{OpenXLA Project}},
  year         = {2026},
  howpublished = {\url{https://openxla.org/xla}},
  note         = {Accessed: 2026-05-08}
}

@software{orbax2024,
  author       = {{Google}},
  title        = {Orbax: Checkpointing and Persistence Utilities for JAX},
  year         = {2024},
  url          = {https://github.com/google/orbax},
  note         = {Accessed: 2026-05-08}
}

@article{martys1996simulation,
  title={Simulation of multicomponent fluids in complex three-dimensional geometries by the lattice Boltzmann method},
  author={Martys, Nicos S and Chen, Hudong},
  journal={Physical review E},
  volume={53},
  number={1},
  pages={743},
  year={1996},
  publisher={APS},
  doi={10.1103/PhysRevE.53.743 }
}

@article{connington2012review,
  title={A review of spurious currents in the lattice Boltzmann method for multiphase flows},
  author={Connington, Kevin and Lee, Taehun},
  journal={Journal of mechanical science and technology},
  volume={26},
  number={12},
  pages={3857--3863},
  year={2012},
  publisher={Springer},
  doi={10.1007/s12206-012-1011-5}
}

@article{ezzatneshan2020evaluation,
  title={Evaluation of equations of state in multiphase lattice Boltzmann method with considering surface wettability effects},
  author={Ezzatneshan, Eslam and Vaseghnia, Hamed},
  journal={Physica A: Statistical Mechanics and its Applications},
  volume={541},
  pages={123258},
  year={2020},
  publisher={Elsevier},
  doi={10.1016/j.physa.2019.123258}
}

@article{zu2013phase,
  title={Phase-field-based lattice Boltzmann model for incompressible binary fluid systems with density and viscosity contrasts},
  author={Zu, YQ and He, Shuisheng},
  journal={Physical Review E—Statistical, Nonlinear, and Soft Matter Physics},
  volume={87},
  number={4},
  pages={043301},
  year={2013},
  publisher={APS},
  doi= {10.1103/PhysRevE.87.043301 }
}

@article{o2018volume,
  title={A volume-of-fluid ghost-cell immersed boundary method for multiphase flows with contact line dynamics},
  author={O’Brien, Adam and Bussmann, Markus},
  journal={Computers \& Fluids},
  volume={165},
  pages={43--53},
  year={2018},
  publisher={Elsevier},
  doi={10.1016/j.compfluid.2018.01.006}
}

@article{xu2017lattice,
  title={Lattice B oltzmann simulation of immiscible two-phase flow with capillary valve effect in porous media},
  author={Xu, Zhiyuan and Liu, Haihu and Valocchi, Albert J},
  journal={Water Resources Research},
  volume={53},
  number={5},
  pages={3770--3790},
  year={2017},
  publisher={Wiley Online Library},
  doi={10.1002/2017WR020373}
}

@article{bhattacharya2026Capillary,
  author       = {Bhattacharya, Malyadeep and Agarwal, Shivam and Dash, Snigdhadyut and Subhedar, Amol},
  title        = {A lattice Boltzmann scheme for contact line capillary force and torque},
  journal      = {SSRN Electronic Journal},
  year         = {2026},
  doi          = {10.2139/ssrn.6584190},
  url          = {http://dx.doi.org/10.2139/ssrn.6584190},
  note         = {Submitted to JCP}
}

@article{aidun2010lattice,
  title={Lattice-Boltzmann method for complex flows},
  author={Aidun, Cyrus K and Clausen, Jonathan R},
  journal={Annual review of fluid mechanics},
  volume={42},
  number={1},
  pages={439--472},
  year={2010},
  publisher={Annual Reviews},
  doi = {10.1146/annurev-fluid-121108-145519}
}

@book{kruger2017lattice,
  title     = {The Lattice Boltzmann Method: Principles and Practice},
  author    = {Kr{\"u}ger, Timm and Kusumaatmaja, Halim and Kuzmin, Alexandr and Shardt, Orest and Silva, Gon{\c c}alo and Viggen, Erlend Magnus},
  year      = {2017},
  publisher = {Springer International Publishing},
  series    = {Graduate Texts in Physics},
  address   = {Cham, Switzerland},
  isbn      = {978-3-319-44647-9, 978-3-319-44649-3},
  doi       = {10.1007/978-3-319-44649-3},
  url       = {https://link.springer.com/book/10.1007/978-3-319-44649-3}
}

@article{li2014contact,
  title={Contact angles in the pseudopotential lattice Boltzmann modeling of wetting},
  author={Li, Qing and Luo, KH and Kang, QJ and Chen, Q},
  journal={Physical Review E},
  volume={90},
  number={5},
  pages={053301},
  year={2014},
  publisher={APS},
  DOI={10.1103/PhysRevE.90.053301 }
}

@article{routh2013drying,
  title={Drying of thin colloidal films},
  author={Routh, Alexander F},
  journal={Reports on Progress in Physics},
  volume={76},
  number={4},
  pages={046603},
  year={2013},
  publisher={IOP Publishing},
    doi={10.1088/0034-4885/76/4/046603}
}

@article{kistler1994teapot,
  title={The teapot effect: sheet-forming flows with deflection, wetting and hysteresis},
  author={Kistler, SF and Scriven, LE},
  journal={Journal of Fluid Mechanics},
  volume={263},
  pages={19--62},
  year={1994},
  publisher={Cambridge University Press},
  DOI={10.1017/S0022112094004027}
}

@article{akai2018wetting,
  title={Wetting boundary condition for the color-gradient lattice Boltzmann method: Validation with analytical and experimental data},
  author={Akai, Takashi and Bijeljic, Branko and Blunt, Martin J},
  journal={Advances in Water Resources},
  volume={116},
  pages={56--66},
  year={2018},
  publisher={Elsevier},
  doi={10.1016/j.advwatres.2018.03.014}
}

@article{yiotis2007lattice,
  title={A lattice Boltzmann study of viscous coupling effects in immiscible two-phase flow in porous media},
  author={Yiotis, Andreas G and Psihogios, John and Kainourgiakis, Michael E and Papaioannou, Aggelos and Stubos, Athanassios K},
  journal={Colloids and Surfaces A: Physicochemical and Engineering Aspects},
  volume={300},
  number={1-2},
  pages={35--49},
  year={2007},
  publisher={Elsevier},
  doi={10.1016/j.colsurfa.2006.12.045}
}

@article{liu2015lattice,
  title={Lattice Boltzmann modeling of contact angle and its hysteresis in two-phase flow with large viscosity difference},
  author={Liu, Haihu and Ju, Yaping and Wang, Ningning and Xi, Guang and Zhang, Yonghao},
  journal={Physical Review E},
  volume={92},
  number={3},
  pages={033306},
  year={2015},
  publisher={APS},
  doi= {10.1103/PhysRevE.92.033306 }
}

@article{fakhari2017improved,
  title={Improved locality of the phase-field lattice-Boltzmann model for immiscible fluids at high density ratios},
  author={Fakhari, Abbas and Mitchell, Travis and Leonardi, Christopher and Bolster, Diogo},
  journal={Physical Review E},
  volume={96},
  number={5},
  pages={053301},
  year={2017},
  publisher={APS},
  doi = {10.1103/PhysRevE.96.053301}
}

@inproceedings{xu2025mitigating,
  title={Mitigating spurious currents at the three phase interfaces in pseudo-potential lattice boltzmann simulations of contact angles via equations of state and forcing schemes},
  author={Xu, Hailin and Wang, Yuxin and Jian, Longzhou},
  booktitle={Journal of Physics: Conference Series},
  volume={3021},
  pages={012086},
  year={2025},
  organization={IOP Publishing},
  doi={10.1088/1742-6596/3021/1/012086}
}

@article{bhattacharya2026hysteresis,
  title={Modeling of contact angle hysteresis for rigid mobile particles: A phase-field lattice Boltzmann approach},
  author={Bhattacharya, Malyadeep and Vaswani, Jovina and Velankar, Sachin S and Subhedar, Amol},
  journal={Physical Review E},
  volume={113},
  number={2},
  pages={025303},
  year={2026},
  publisher={APS},
  doi= {10.1103/59kb-q12s}
}

@article{liu2015diffuse,
  title={A diffuse-interface immersed-boundary method for two-dimensional simulation of flows with moving contact lines on curved substrates},
  author={Liu, Hao-Ran and Ding, Hang},
  journal={Journal of Computational Physics},
  volume={294},
  pages={484--502},
  year={2015},
  publisher={Elsevier},
  doi={10.1016/j.jcp.2015.03.059}
 
}

@article{fakhari2017diffuse,
  title={Diffuse interface modeling of three-phase contact line dynamics on curved boundaries: A lattice Boltzmann model for large density and viscosity ratios},
  author={Fakhari, Abbas and Bolster, Diogo},
  journal={Journal of Computational Physics},
  volume={334},
  pages={620--638},
  year={2017},
  publisher={Elsevier},
  doi={10.1016/j.jcp.2017.01.025}
}

@article{lee2008wall,
  title={Wall boundary conditions in the lattice Boltzmann equation method for nonideal gases},
  author={Lee, Taehun and Liu, Lin},
  journal={Physical Review E—Statistical, Nonlinear, and Soft Matter Physics},
  volume={78},
  number={1},
  pages={017702},
  year={2008},
  publisher={APS},
  DOI= {10.1103/PhysRevE.78.017702 }
}

@software{jax2018github,
  author = {James Bradbury and Roy Frostig and Peter Hawkins and Matthew James Johnson and Chris Leary and Dougal Maclaurin and George Necula and Adam Paszke and Jake Vander{P}las and Skye Wanderman-{M}ilne and Qiao Zhang},
  title = {{JAX}: composable transformations of {P}ython+{N}um{P}y programs},
  url = {http://github.com/google/jax},
  version = {0.3.13},
  year = {2018},
}

@article{zhang2023simplified,
  title={Simplified wetting boundary scheme in phase-field lattice Boltzmann model for wetting phenomena on curved boundaries},
  author={Zhang, Shengyuan and Tang, Jun and Wu, Huiying},
  journal={Physical Review E},
  volume={108},
  number={2},
  pages={025303},
  year={2023},
  publisher={APS},
  DOI= {10.1103/PhysRevE.108.025303 }
}

@article{ding2007wetting,
  title={Wetting condition in diffuse interface simulations of contact line motion},
  author={Ding, Hang and Spelt, Peter DM},
  journal={Physical Review E—Statistical, Nonlinear, and Soft Matter Physics},
  volume={75},
  number={4},
  pages={046708},
  year={2007},
  publisher={APS},
  doi={10.1103/PhysRevE.75.046708 }
}

@article{liang2019lattice,
  title={Lattice Boltzmann method for contact-line motion of binary fluids with high density ratio},
  author={Liang, Hong and Liu, Haihu and Chai, Zhenhua and Shi, Baochang},
  journal={Physical Review E},
  volume={99},
  number={6},
  pages={063306},
  year={2019},
  publisher={APS},
  DOI= {10.1103/PhysRevE.99.063306 }
}

@article{bergeron2000controlling,
  title={Controlling droplet deposition with polymer additives},
  author={Bergeron, Vance and Bonn, Daniel and Martin, Jean Yves and Vovelle, Louis},
  journal={Nature},
  volume={405},
  number={6788},
  pages={772--775},
  year={2000},
  publisher={Nature Publishing Group UK London},
  doi={10.1038/35015525}
}

@article{song2022droplet,
  title={Droplet spreading characteristics on ultra-slippery solid hydrophilic surfaces with ultra-low contact angle hysteresis},
  author={Song, Yajie and Wang, Qi and Ying, Yushan and You, Zhuo and Wang, Songbai and Chun, Jiang and Ma, Xuehu and Wen, Rongfu},
  journal={Coatings},
  volume={12},
  number={6},
  pages={755},
  year={2022},
  publisher={MDPI},
  doi={10.3390/coatings12060755}
}

@article{shams2018numerical,
  title={A numerical model of two-phase flow at the micro-scale using the volume-of-fluid method},
  author={Shams, Mosayeb and Raeini, Ali Q and Blunt, Martin J and Bijeljic, Branko},
  journal={Journal of Computational Physics},
  volume={357},
  pages={159--182},
  year={2018},
  publisher={Elsevier},
  doi = {10.1016/j.jcp.2017.12.027}
}

@article{leclaire2017generalized,
  title={Generalized three-dimensional lattice Boltzmann color-gradient method for immiscible two-phase pore-scale imbibition and drainage in porous media},
  author={Leclaire, S{\'e}bastien and Parmigiani, Andrea and Malaspinas, Orestis and Chopard, Bastien and Latt, Jonas},
  journal={Physical Review E},
  volume={95},
  number={3},
  pages={033306},
  year={2017},
  publisher={APS},
  doi= {10.1103/PhysRevE.95.033306 }
}

@article{subhedar2022color,
  title={Color-gradient lattice Boltzmann model for immiscible fluids with density contrast},
  author={Subhedar, A},
  journal={Physical Review E},
  volume={106},
  number={4},
  pages={045308},
  year={2022},
  publisher={APS},
  DOI= {10.1103/PhysRevE.106.045308 }
}

@article{younes2022phase,
  title={Phase-field lattice Boltzmann model for liquid bridges and coalescence in wet granular media},
  author={Younes, Nabil and Benseghier, Z and Millet, O and Wautier, A and Nicot, F and Wan, R},
  journal={Powder Technology},
  volume={411},
  pages={117942},
  year={2022},
  publisher={Elsevier},
  doi={10.1016/j.powtec.2022.117942}
}

@article{sbragaglia2007diffOperators,
  title={Generalized lattice Boltzmann method with multirange pseudopotential},
  author={Sbragaglia, MRLSK and Benzi, Roberto and Biferale, Luca and Succi, Sauro and Sugiyama, Kazu and Toschi, Federico},
  journal={Physical Review E—Statistical, Nonlinear, and Soft Matter Physics},
  volume={75},
  number={2},
  pages={026702},
  year={2007},
  publisher={APS},
  doi={10.1103/PhysRevE.75.026702}
}

@article{kotnala2017microfluidic,
  title={Microfluidic-based high-throughput optical trapping of nanoparticles},
  author={Kotnala, Abhay and Zheng, Yi and Fu, Jianping and Cheng, Wei},
  journal={Lab on a Chip},
  volume={17},
  number={12},
  pages={2125--2134},
  year={2017},
  publisher={Royal Society of Chemistry},
  doi={10.1039/C7LC00286F}
}

@article{subhedar2020interface,
  title={Interface tracking characteristics of color-gradient lattice Boltzmann model for immiscible fluids},
  author={Subhedar, A and Reiter, A and Selzer, M and Varnik, F and Nestler, B},
  journal={Physical Review E},
  volume={101},
  number={1},
  pages={013313},
  year={2020},
  publisher={APS},
  doi={10.1103/PhysRevE.101.013313}
}

@article{wang2024wetting,
  title={Wetting boundary condition for three-dimensional curved geometries in lattice Boltzmann color-gradient model},
  author={Wang, Ningning and Kuang, Tie and Liu, Yong and Yin, Zhilin and Liu, Haihu},
  journal={Physics of Fluids},
  volume={36},
  number={3},
  pages = {032133},
  year={2024},
  publisher={AIP Publishing},
  doi={10.1063/5.0200478}
}

@article{parker1989multiphase,
  title={Multiphase flow and transport in porous media},
  author={Parker, JC},
  journal={Reviews of Geophysics},
  volume={27},
  number={3},
  pages={311--328},
  year={1989},
  publisher={Wiley Online Library},
  doi={10.1029/RG027i003p00311}
}

@article{guzman2022broad,
  title={A broad perspective to particle-laden fluid interfaces systems: From chemically homogeneous particles to active colloids},
  author={Guzm{\'a}n, Eduardo and Martinez-Pedrero, Fernando and Calero, Carles and Maestro, Armando and Ortega, Francisco and Rubio, Ram{\'o}n G},
  journal={Advances in colloid and interface science},
  volume={302},
  pages={102620},
  year={2022},
  publisher={Elsevier},
  doi={10.1016/j.cis.2022.102620}
}

@article{bakshi2007investigations,
  title={Investigations on the impact of a drop onto a small spherical target},
  author={Bakshi, Shamit and Roisman, Ilia V and Tropea, Cam},
  journal={Physics of fluids},
  volume={19},
  number={3},
  pages = {032102},
  year={2007},
  publisher={AIP Publishing},
  doi={10.1063/1.2716065}
}

@article{yarin1995impact,
  title={Impact of drops on solid surfaces: self-similar capillary waves, and splashing as a new type of kinematic discontinuity},
  author={Yarin, Alexander L and Weiss, Daniel A},
  journal={Journal of Fluid Mechanics},
  volume={283},
  pages={141--173},
  year={1995},
  publisher={Cambridge University Press},
  doi={10.1017/S0022112095002266}
}

@article{yao2022multiphase,
  title = {Multiphase curved boundary condition in lattice Boltzmann method},
  author = {Yao, Yichen and Liu, Yangsha and Zhong, Xingguo and Wen, Binghai},
  journal = {Phys. Rev. E},
  volume = {106},
  issue = {1},
  pages = {015307},
  numpages = {12},
  year = {2022},
  month = {Jul},
  publisher = {American Physical Society},
  doi = {10.1103/PhysRevE.106.015307}
}

@article{fei2024pore,
  title={Pore-scale study on shear rheology of wet granular materials},
  author={Fei, Linlin and He, Ya-Ling and Derome, Dominique and Carmeliet, Jan},
  journal={Physics of Fluids},
  volume={36},
  number={11},
  year={2024},
  publisher={AIP Publishing},
  url = {https://doi.org/10.1063/5.0243150}
}

@article{tao2016investigation,
title = {An investigation on momentum exchange methods and refilling algorithms for lattice Boltzmann simulation of particulate flows},
journal = {Computers \& Fluids},
volume = {133},
pages = {1-14},
year = {2016},
issn = {0045-7930},
doi = {https://doi.org/10.1016/j.compfluid.2016.04.009},
author = {Shi Tao and Junjie Hu and Zhaoli Guo}
}

@article{he2022improved,
    author = {He, Qiang and Huang, Weifeng and Yin, Yuan and Hu, Yang and Li, Yanwen and Li, Decai},
    title = {An improved lattice Boltzmann model for fluid–fluid–solid flows with high viscosity ratio},
    journal = {Physics of Fluids},
    volume = {34},
    number = {9},
    pages = {093322},
    year = {2022},
    month = {09},
    issn = {1070-6631},
    doi = {10.1063/5.0107431},
    url = {https://doi.org/10.1063/5.0107431}
}
\bibliographystyle{apsrev4-2}
\end{document}